\documentclass[useAMS,usenatbib]{mn2e}
\usepackage{graphicx}
\usepackage{rotating}
\usepackage{longtable}
\usepackage{lscape}
\usepackage{amssymb,amsmath}
\begin{document}

\title[Low z QSO properties]
{Low redshift  quasars in the SDSS Stripe 82. The host galaxies. }
% Version : 7 February  2014 to the Editor

\author[Falomo  et al.]{
	R. Falomo$^{1}$\thanks{E--mail: {\tt renato.falomo@oapd.inaf.it}}, 
        D. Bettoni$^{1}$,
	K. ~Karhunen$^{2}$, 
	J.~K.~Kotilainen$^{3}$, and 
	M. Uslenghi$^{4}$ \\
       	$^{1}$ INAF -- Osservatorio Astronomico di Padova, Vicolo dell'Osservatorio 5, I-35122 Padova (PD), Italy\\
       	$^{3}$ Finnish Centre for Astronomy with ESO (FINCA), University of Turku, V\"ais\"al\"antie 20, FI-21500 Piikki\"o, Finland\\
       	$^{2}$ Tuorla Observatory, Department of Physics and Astronomy, University of Turku, FI-21500 Piikkio, Finland.\\
       	$^{4}$ INAF-IASF via E. Bassini 15, 20133 - Milano, Italy
	}

\maketitle

\label{firstpage}

\begin{abstract}

We present a photometrical and morphological study of the properties of low redshift (z $<$ 0.5)
quasars based on a large and homogeneous dataset of objects derived from the Sloan Digital Sky Survey 
(DR7). This study over number by a factor $\sim$ 5 any other previous study of QSO host galaxies at low
redshift undertaken either on ground or on space surveys. We used $\sim$ 400 quasars that were imaged
in the SDSS Stripe82 that is up to 2 mag deeper than standard Sloan images. For these quasars we
undertake a study of the host galaxies and of their environments. In this paper we report the results
for the quasar hosts.

We are able to detect the host galaxy for more than  3/4 of the whole dataset and characterise the properties of their
hosts. We found that QSO hosts are dominated by luminous galaxies of absolute magnitude M*-3 $<$ M(R)
$<$ M*. For the unresolved objects we computed a upper limit to the host luminosity. For each well resolved
quasar we are also able to characterise the morphology of the host galaxy that turn out to be more complex than what
found in previous studies. QSO are hosted in a variety of galaxies from pure ellipticals to complex/composite
morphologies that combine spheroids, disk, lens and halo.

The black hole mass of the quasar, estimated from the spectral properties of the nuclei, are poorly correlated with the total 
luminosity of the host galaxy. However, taking into account only the bulge component we found a significant 
correlation between the BH mass and the bulge luminosity of the host.
 
\end{abstract}

\bigskip

\section{Introduction}

Accretion onto a supermassive black hole (SMBH) is the main mechanism that sustains the powerful activity of
active galactic nuclei but may also represent a common phase in the evolution of normal galaxies. A number of
fundamental question about the formation of the QSO phenomenon like the fuelling and triggering mechanisms are
strictly related to the immediate environments of the active nucleus and in particular to its host galaxy 
\citep{merloni10}. 
SMBHs may well have a period of maximum growth (maximum nuclear luminosity) contemporaneous with the bulk of the
initial star formation in the bulge of galaxies. Studies of the co-evolution of SMBH and their host spheroids
are therefore obviously critical to understanding how and when galaxies in the local Universe formed and
evolved. The last ten years have yielded considerable progress in characterising AGN host galaxies. At variance
with inactive galaxies their study is often hampered by the presence of the luminous central source that
outshines the light of the host galaxy. A problem that becomes more serious for high luminosity AGNs and for
sources at high redshift.

In spite of these limitations the characterization of the properties of the host galaxies offers the unique
opportunity to investigate the link between the central black hole mass and its host galaxy at moderate to high
redshift and to trace the possible co-evolution at different cosmic epochs. 
This is because for broad line AGN like
quasars it is possible to estimate the mass of the central BH using kinematic arguments that are not directly
dependent on the host galaxies properties.

% --- OVERVIEW OF QSO HOST GALAXIES ... PREVIOUS WORKS

Both ground-based and HST studies have shown that virtually all luminous low redshift (z$<$0.5) quasars reside
in massive, spheroid-dominated host galaxies, whereas at lower luminosities quasars can also be found in early-
type spiral hosts (e.g. Bahcall et al. 1997; Dunlop et al. 2003; Pagani et al. 2003; Floyd et al. 2004;
Jahnke et al. 2004). This is in good agreement with the BH -- bulge relationship in inactive galaxies (e.g.
Gultekin et al. 2009), since very massive BHs power luminous quasars. Only a small fraction of the host galaxies
($\sim$15\% ) are found in merger systems but it is difficult to determine clear merger signatures from
morphology alone. At low redshifts a major contribution to the properties of quasar host galaxies has been
provided by images from the Hubble Space Telescope (HST). The improved spatial resolution has allowed the
characterization of the structure and the detailed morphology of the host galaxies \citep{bahcall97, kukula01,
ridgway01,dunlop03,peng06,zakamska06}. It turned out that QSO are hosted in luminous galaxies that are often
dominated by the spheroidal component.

At high redshift (z $>$1) HST observations of quasar host galaxies (e.g. \cite{peng06,floyd13} and references
therein) have been complemented by significant contributions from 8-m class ground-based telescopes under
superb seeing conditions \citep{kotilainen07,kotilainen09} and/or with adaptive optics  \citep{falomo08}.
Comparison of host galaxies of AGN at high and low redshift constrain host galaxy evolution, as compared with 
the evolution of normal (inactive) galaxies. 

%--- WHY A NEW STUDY ON QSO HOST GALAXIES

Most of the old studies of quasar host considered few tens of objects therefore in order to derive a picture of
the host properties at various redshift one should combine many different samples often obtained with different
telescopes and filters. Observations carried out by HST are certainly more homogeneous (although different
filters were used) and allow to investigate  a somewhat large sample based on high quality data. Nevertheless
the size of these samples remain relatively small For instance in the range 0.25 $ < z < $ 0.5 about 50 QSO were
imaged by HST (see references above).

In order to explore a significantly larger dataset of QSO one should refer to large surveys that include both
imaging and spectroscopic data. In this respect one of the most productive recent surveys is the Sloan Digital
Sky Survey that allowed to find  105783 quasars \citep{schneider2010} from (DR-7).  Standard SDSS images are,
however, too shallow and the faint nebulosity around the nucleus of quasars is not  detected. This problem
has been overcome in the case of the special sky region mapped by SDSS for the SDSS Legacy Survey.

The central stripe in the South Galactic Gap, namely the Stripe82 \citep{annis2011} is a stripe along the Celestial Equator in the
Southern Galactic Cap. It is 2.5$^{\deg}$ wide and covers -50$^{\deg}\leq$RA$\leq$+ 60$^{\deg}$, so its total area is 275deg$^2$.
Stripe 82 was imaged by the SDSS multiple times, these data were taken in 2004 only under optimal seeing, sky brightness, and
photometric conditions (i.e., the conditions required for imaging in the main Legacy Survey; York et al. (2000)). In 2005-2007, 219
additional imaging runs were taken on Stripe 82 as part of the SDSS supernova survey \citep{frieman08}, designed to discover Type
Ia supernovae at 0.1$< z <$0.4. The total number of images reaches $\sim$100 for the S strip and $\sim$ 80 for the N strip. The
final frames were obtained by co-adding selected fields in r-band, with seeing (as derived from 2D gaussian fit of stars and
provided by SDSS pipeline) better than 2", sky brightness $\leq$19.5 $mag/arcsec^2$ and less than 0.2 mag of extinction. In this
area there are 12434 quasars. 

Recently \cite{matsuoka14} analyzed the stellar properties of about 800 galaxies hosting
optically luminous, unobscured quasars at z$<$ 0.6 using Stripe82 images. They focused on the color of the host galaxies 
and found that the quasar hosts are very blue and almost
absent on the red sequence with a marked different distribution from that of normal (inactive) galaxies. 

For our study we selected QSO with redshift less than 0.5 for which the stripe 82 images allow us also to
study the QSO galaxy environments. We adopt the concordance cosmology with H$_0$ = 70 km s$^{-1}$ Mpc$^{-1}$, $\Omega_m$ = 0.3 and
$\Omega_\Lambda$ = 0.7.

 In this first paper of a series we focus on the properties of quasar hosts and their relationship with 
the central BH mass in the explored redshift range. In forthcoming papers we investigate the galaxy environments 
\citep{karhunen13}  and galaxy peculiarities \citep{bettoni14}. A preliminary account of these results was presented in 
\cite{kotilainen13}.

\section{The low z QSO sample}

\begin{figure*}
\centering
\includegraphics[width=2.0\columnwidth]{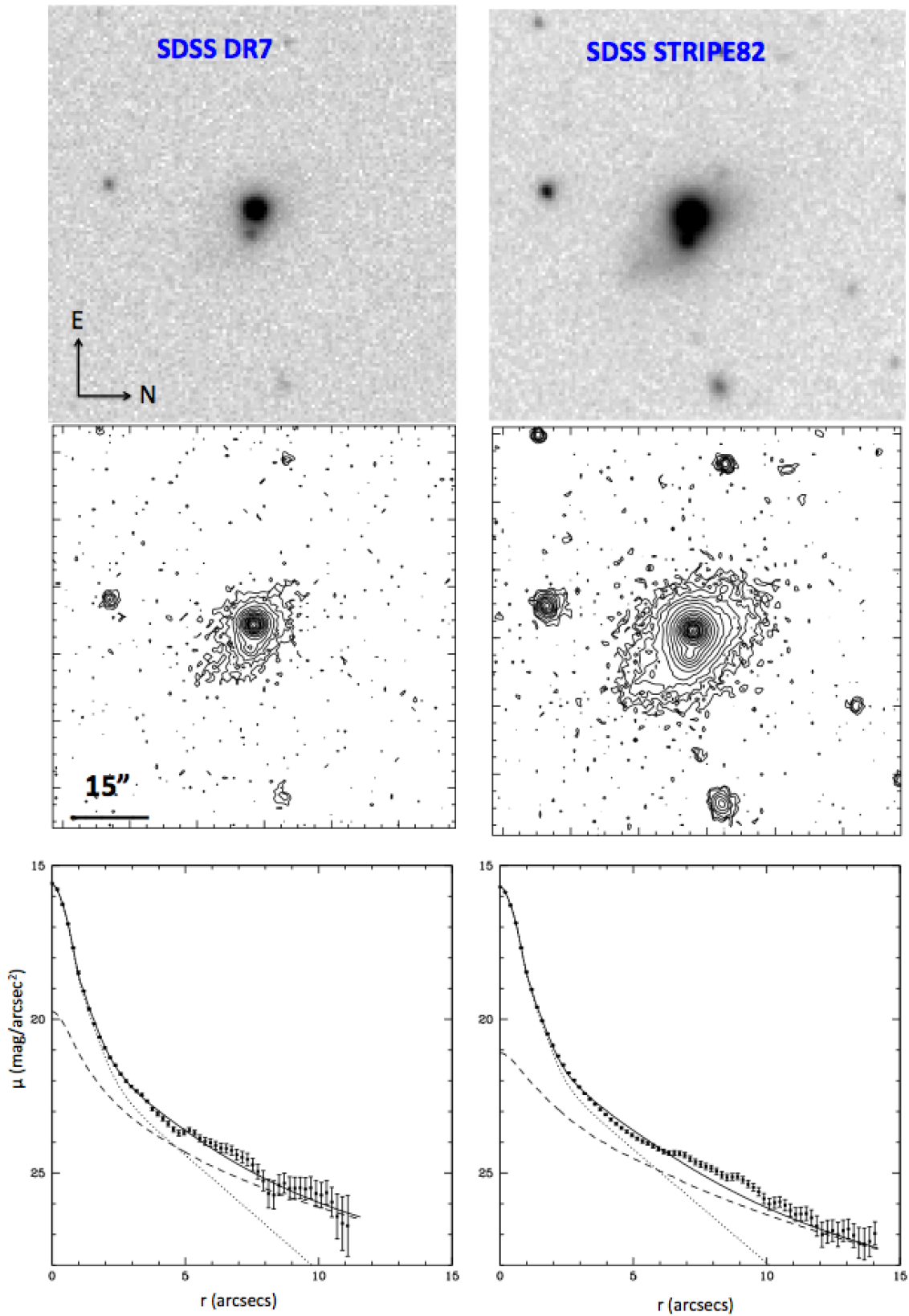}
\caption{ Example of QSO in the sample:  Left panels show the SDSS DR7 data; Right panels  the corresponding data from Stripe 82 (
image resulting combining 35 individual images of 54 sec). Top panels yield the grey scale images in the i band; central panels give
contour plots of the region and in the bottom panels we show the luminosity radial profiles together with the AIDA fit (see text for
details) } \label{fig:dr7_sdss82}
\end{figure*}

To derive the sample of low redshift quasars we used the fifth release of the SDSS Quasar Catalog 
 \citep{schneider2010} that is based on the SDSS-DR7 data release \citep{abazajian09}. It consists of
 QSO that have a highly reliable redshift measurement and are fainter than $i\sim15.0$, that have an
 absolute magnitude $M_i<$-22, at least one emission line with FWHM$>$1000 km/ sec, or have complex/
 interesting absorption lines. This catalog contain $\sim$106.000 spectroscopically confirmed quasars.
 Our analysis is done only in the region of sky covered by the stripe82 data, these images go deeper of
 about $\sim$2  magnitudes with respect to the usual Sloan data and make possible the study of the QSO
 hosts (see example in Fig. \ref{fig:dr7_sdss82}).

We apply two main constraints on the sample. First, we avoid objects that are closer than 0.2 deg to
the edges of the Stripe82 \citep{annis2011,abazajian09}. Second, we choose an upper redshift
limit z=0.5 in order to be able to resolve the quasar host for the large majority of the sample.

\begin{figure*}
\centering
\includegraphics[width=1.9\columnwidth]{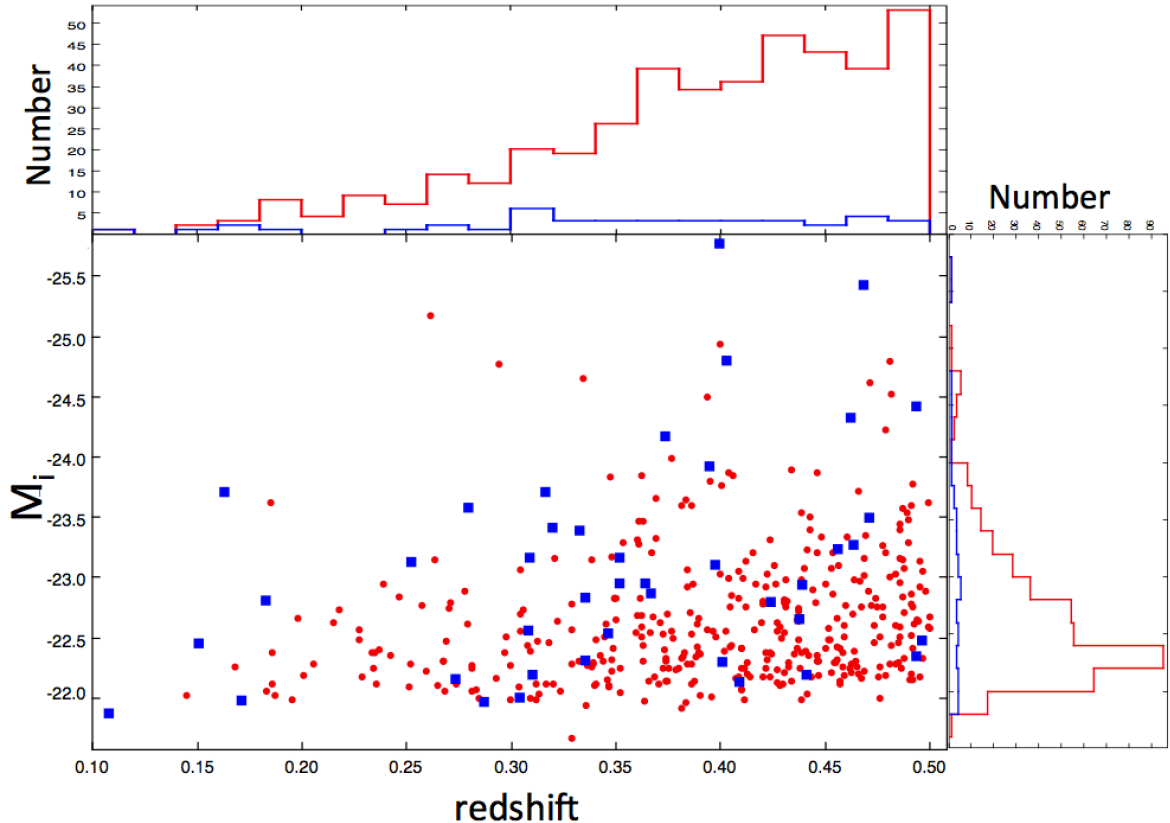}
\caption{The QSO sample in the $M_i$ redshift plane. Red filled circles are radio quiet while 
blue filled squares are  radio loud QSOs. 
Top and right panels show the distributions of the redshift and luminosity 
for radio quiet (red) and radio loud (blue) objects.
}
\label{fig:qsosample}
\end{figure*}

To satisfy all these reasons  we therefore select all the QSOs in the range of redshift $0.1<z<0.5$ and in the Stripe82
region i.e. $1.0<DEC<-1.0$, $0<RA<59.8$ and $300.2<RA<360$. This gives a total of 416 QSOs. In this sample we are dominated by
radio quiet quasars only 24 are radio loud (about 5\%). In Fig. \ref{fig:qsosample} we report the distribution of QSO in the
plane z-$M_i$ . The mean redshift of the sample is $<z>$ = $0.39\pm0.08$ (median $0.41\pm0.06$ ) and the average absolute
magnitude is : $<M_i>$ = $-22.68\pm0.61$ (median $-22.52\pm0.35$ ).

In Table \ref{tab1} we report the main data for the QSO in the sample. In column (1) id number, in column (2) the SDSS
identification in columns (3) and (4) the coordinates, in column (5) the redshift, in column (6) the i band psf magnitude from
SDSS-DR7, in column (7) the absolute i band magnitude, in column (8) the number of exposures for each co added frame, in
column (9) the i band extinction and finally in column (10) the measured seeing on the co-added images. The images used have
an average seeing, as given by 2D gaussian fit of stars in the frame from SDSS, of 1.20$\pm$0.09 arcsec, with a
minimum of 1.01 and a maximum of 1.47 arcsec.

% ---- FIG 1 --------------------------

% ---- FIG 2 --------------------------

\section{Image analysis}

We have retrieved all images of the selected QSO from SDSS Stripe 82 dataset  \citep{annis2011} in the i band. 
This corresponds to observe in the R filter at rest frame at the average redshift of the dataset. 
In order to derive the properties of the galaxies hosting the QSO we performed a 2D fit of the
image of the QSO assuming it is the superposition of two components. The nucleus in the center
and the surrounding nebulosity. The first is described by the local Point Spread Function of the
image while for the second component we assumed  a galaxy model described by a Sersic law
convolved with the proper PSF. The analysis of these images was performed using the Astronomical
Image Decomposition Analysis (AIDA, \cite{uslenghi08}) that was used in our previous studies  
of QSO host galaxies \citep{falomo08,decarli2012,kotilainen07,kotilainen09}.

The most critical aspect of the image decomposition is the determination of a suitable PSF. In
the case of SDSS images the field of view is large enough that there are always many stars in the
 co-added SDSS image containing the target to properly derive the PSF. 
To derive the most suitable PSF of each field we have selected
a number of stars (between 5 and 15) in the field that are distributed around the target.
Selection of these PSF stars was based on various parameters as their magnitude, FWHM,
ellipticity and presence of close companions. 
In particular, the selection of the stars was done
according to the following criteria: the stars are not saturated; the stars are as close to
the target as possible (while avoiding the fit region of the target); the stars are as
uniformly distributed around the target as possible; the stars are sufficiently isolated
(i.e., they have no close companions); the stars cover a suitably wide range of magnitude in
order to assure that the extended halo of the PSF is well characterised.

We then define a radius to compute the PSF model
and a ring around each star where to compute the sky background. All extra sources that were
found inside these areas were masked out with an automated procedure. The PSF model was then
obtained  from the simultaneous fit of all selected stars using a multi function 2D model composed
by 3 gaussians and one exponential function. 

\begin{figure*}
\centering
\includegraphics[width=1.7\columnwidth]{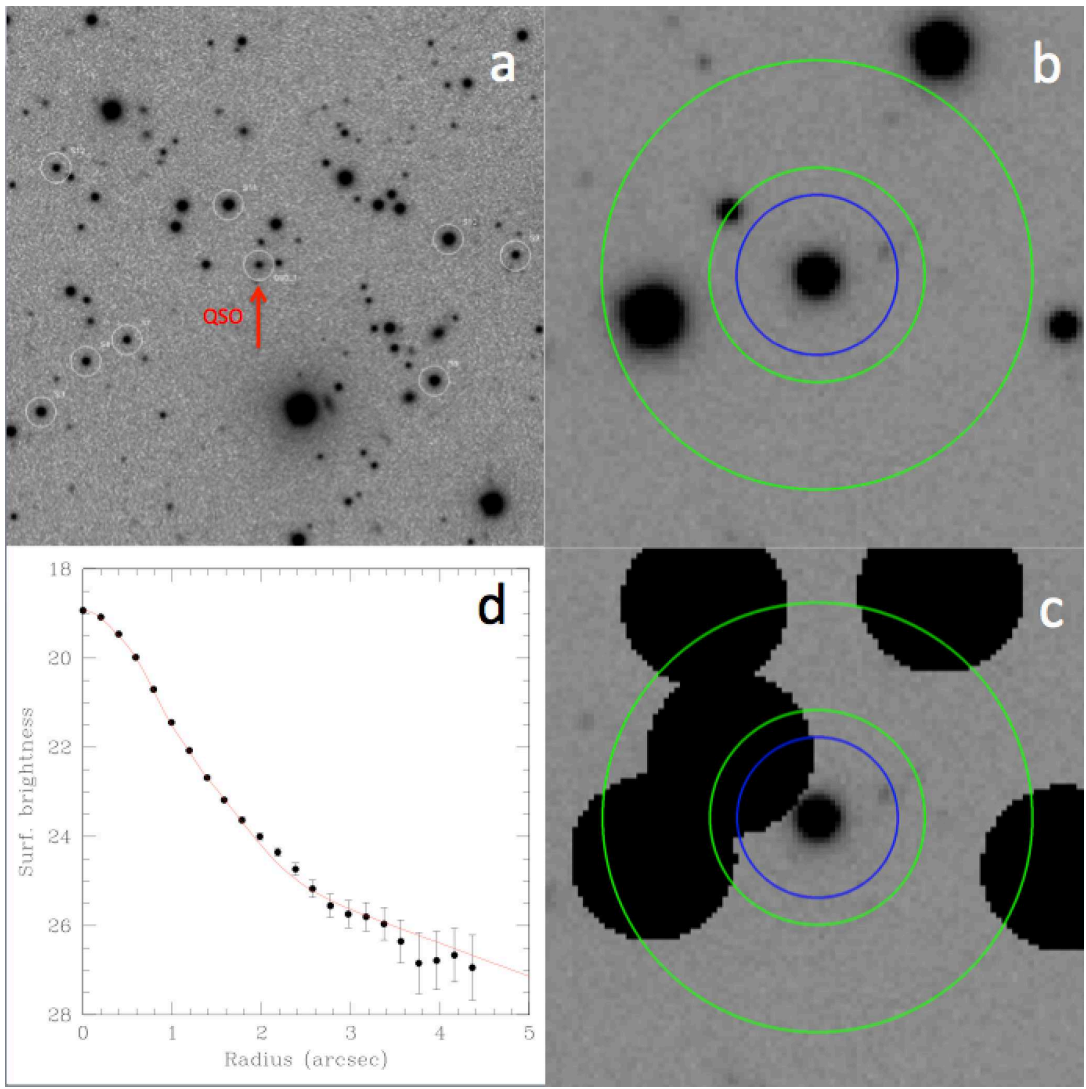}
\caption{Modelling of the PSF: a) Selection of stars around the target; b) example of definition of fit area 
(inner blue circle) and the background region (encompassed by the two green circles) around one selected star; 
c) masked areas to avoid spurious sources; 
d) example of the model fit to the radial brightness profile of one selected star. }
\label{fig:psf}
\end{figure*}

\begin{figure*}
\centering 
\includegraphics[bb=40 500 530 679 ,width=1.9\columnwidth]{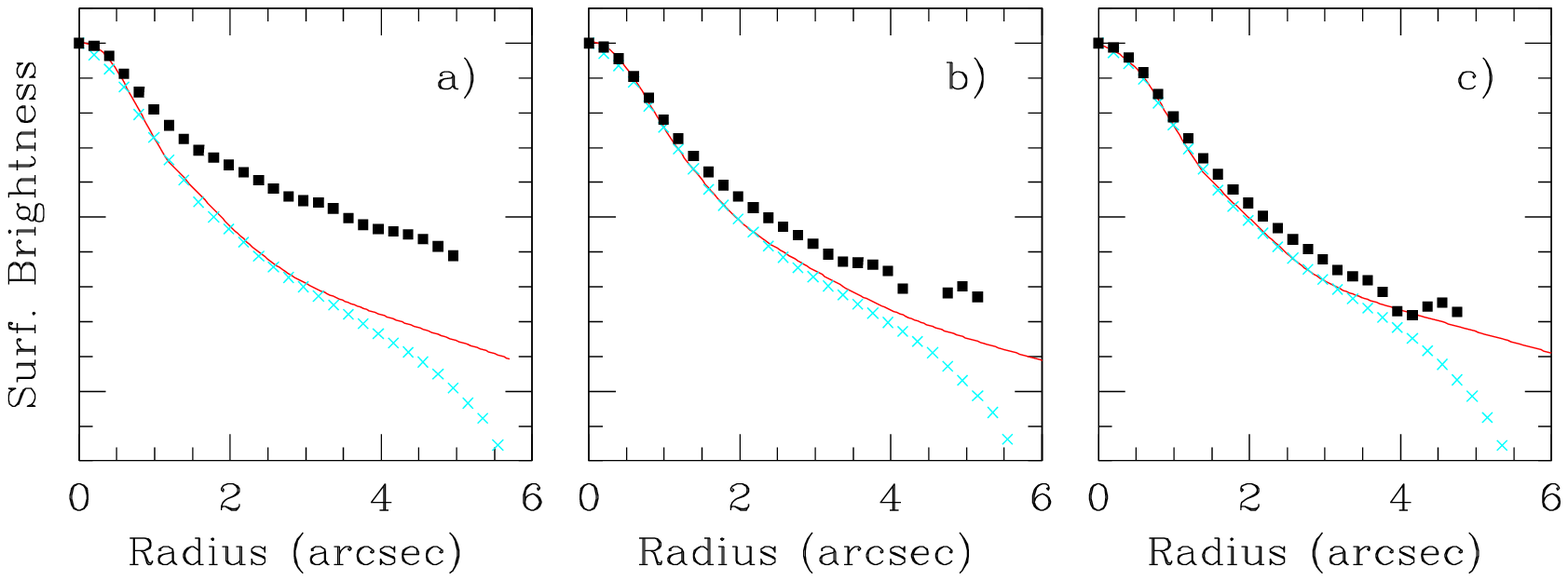}
\includegraphics[bb=40 500 530 679 ,width=1.9\columnwidth]{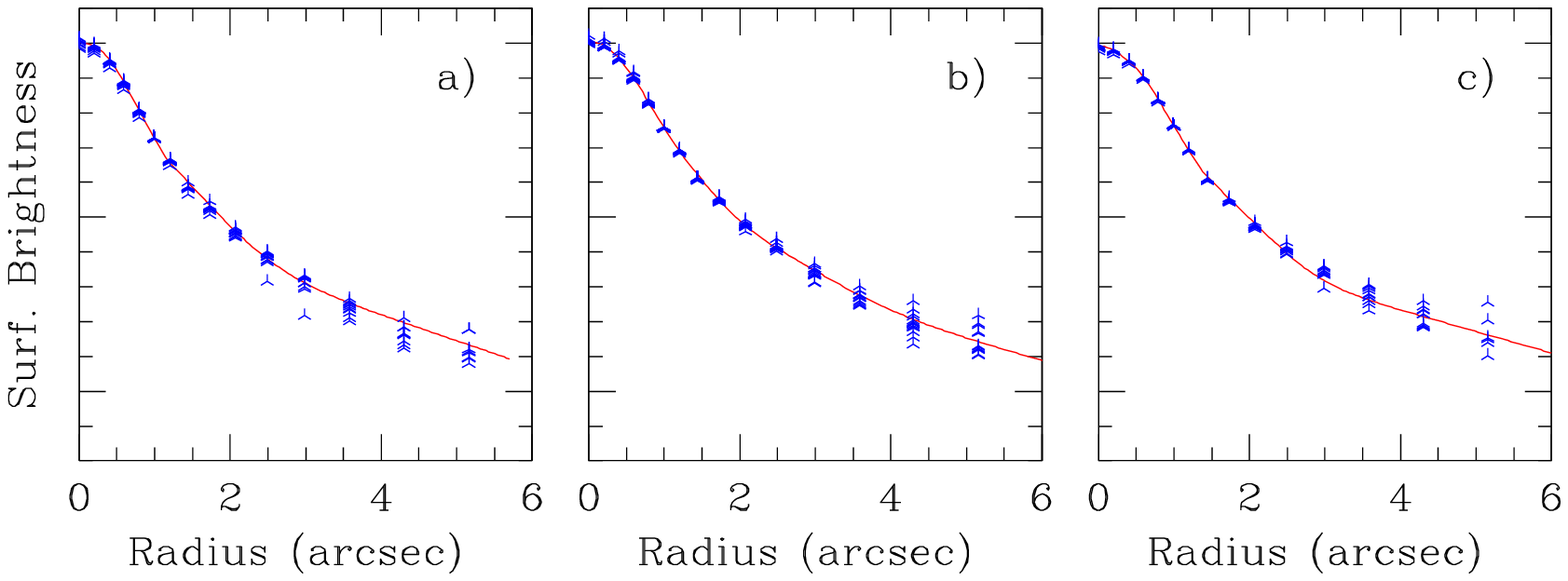}
\caption{ {\it Top panels}: Example of comparison between our PSF model (solid red line) and the PSF 
derived from SDSS archive (psField; light blue crosses) with the average radial brightness profile of three QSO (black 
filled squares). 
There is an excellent agreement for the two PSF until $\sim$ 3 arcsec from the center but beyond this radius 
the SDSS PSF systematically underestimate the flux from the wing of the psf. 
Panels a) (seeing = 1.1  arcsec) and b)  (seeing = 1.22) show  the average radial brightness distribution 
of well and marginally resolved objects, respectively,  
while the one in panel c) is unresolved.
{\it Bottom panels}: Example of comparison between the average radial brightness profile of the adopted PSF model (solid red line) and that of the stars used to derive the PSF (blue points). }

\label{fig:psfconf}
\end{figure*}

In Figure \ref{fig:psf} we show an example of the procedure adopted to derive the  PSF model.
It is worth to note that the PSF provided by the SDSS pipeline (using source psField ) are not 
adequate for this study. This is due to a systematic underestimate of the wings of the SDSS PSF. 
The agreement with our PSF model is excellent up to $\sim$ 3 arcsec from the center of the star but 
then a significant deviation is present (see Figure \ref{fig:psfconf}). Using the SDSS psf 
for the QSO decomposition will result in a systematic overestimate of the host galaxy luminosity and in some cases 
to false detection of the host galaxy signal (see example in Figure \ref{fig:psfconf} panel c) )

% ---- FIG 5 ----------------------------- figure with PSF ---
%\clearpage
% comparison of PSF - AIDA vs SDSS psField

The second step of the analysis is to fit each quasar both with a scaled PSF and
with a 2 components model (point source plus a galaxy). 
 The best fit was obtained  adopting a  model
for the errors that include a constant term to represent the read out noise of the
detector, a term representing the statistical noise due to the effective counts and an
additional term that take into account the possible residual noise due to fixed pattern
noise. 
 For the coadded images we assume a readout noise of 9.5 e$^-$ and an average gain of 3.8 e$^-$/ADU.
The term for the statistical noise is given by 
the coefficient 1/$\sqrt{GAIN \times NEXP}$ that multiply the root square of the counts.
For the residual pattern noise we assumed 2\% value.

% --- figure with FIT---FIG 5 
%\clearpage
\begin{figure*}
\centering
\includegraphics[width=2.0\columnwidth]{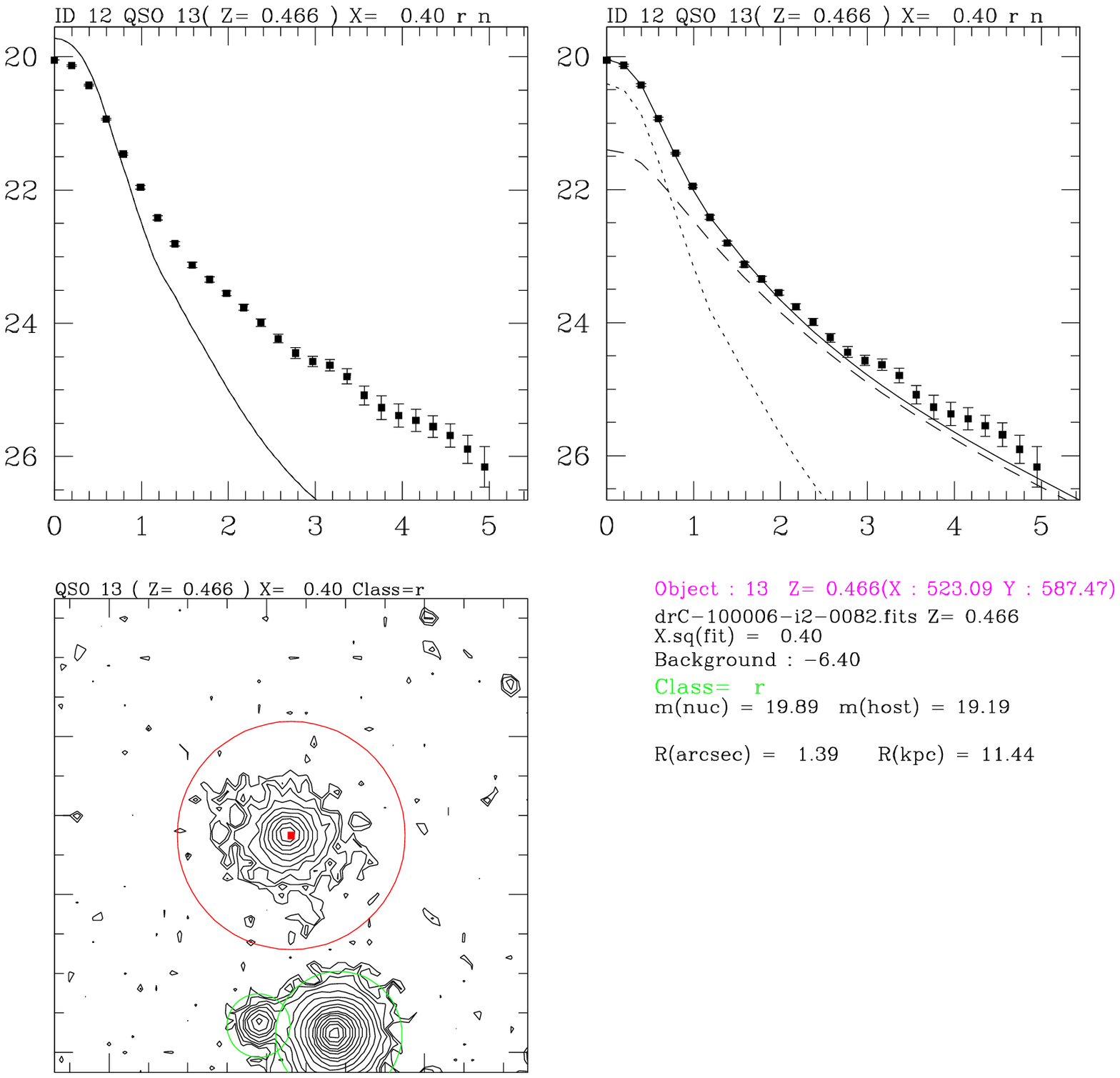}
\caption{Example of AIDA results for a resolved QSO. Bottom-left panel shows the contour plot of the target, 
the area of the fit (red circle) and the masked out objects in the field (green circles).
Top-left panel shows the average radial brightness profile of the target and the best fit by a PSF model.
Top-right panel: same as left panel but for  the best fit of the target by nucleus (PSF) 
and host galaxy (Sersic law model).
Solid line (best fit); dotted line PSF model; dashed line galaxy model convolved with psf}.

\label{fig:aida}
\end{figure*}
\begin{figure}
\centering
\includegraphics[bb=70 150 530 660,width=0.95\columnwidth]{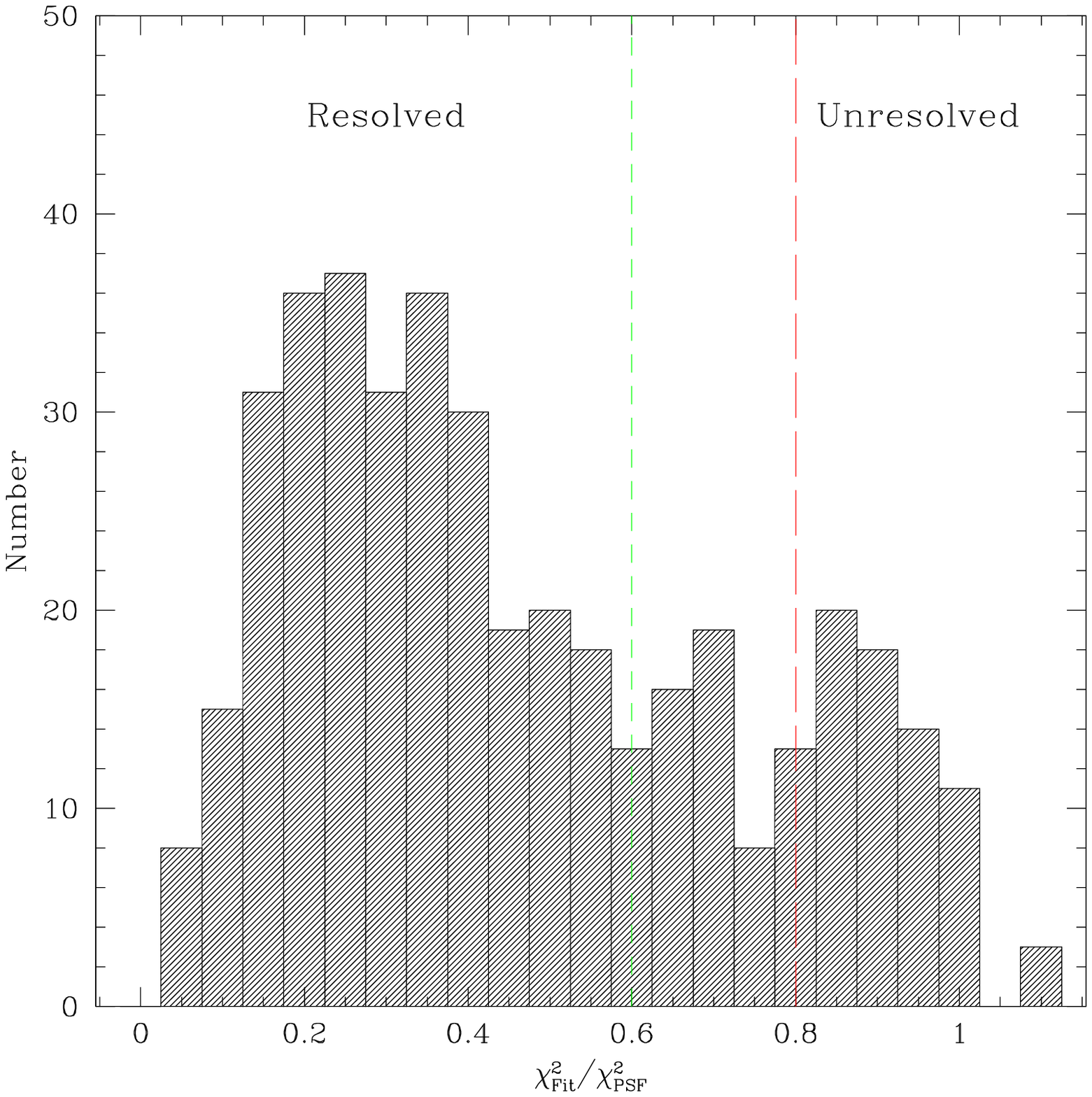}
\caption{Distribution of the $\chi^2$ ratio of the fit (PSF+galaxy) with respect to 
the one with only the PSF). The two vertical lines define the regions of our classification for objects 
resolved ($\chi^2_{Fit}$/$\chi^2_{PSF}$  $<$ 0.6), marginally resolved and unresolved ($\chi^2_{Fit}$/$\chi^2_{PSF}$  $>$ 0.8).}
\label{fig:chi2ratio}
\end{figure}

In Figure \ref{fig:aida} we show an example of the adopted procedure.
In order to distinguish between resolved and unresolved objects we compared the $\chi^2$ of the two fits and
in addition we then inspect all the fits to further check the results. This allows us to
produce a clean list of $\sim$ 350  resolved quasars by removing 7 objects ($\sim$1\%
of the entire sample) that were contaminated by very bright sources in the field or 
defects in the image close to the targets. For 
60 objects ($\sim$13\% of the entire sample) the QSO  were unresolved and the 
 fit with only the psf was indistinguishable from the fit with the
psf + Sersic function (see Figure \ref{fig:chi2ratio}). 
The unresolved objects are mainly objects at relatively high redshift 
(42 out of 60 unresolved quasars are at z $>$ 0.4) and bright  nuclei.

%\clearpage
% ---- FIG -- 5 -----------------------------
%\clearpage

For the unresolved objects we evaluated the bright limit of the host galaxy by adding 
the flux of a galaxy to the observed object until the $\chi^2$ of fit to these data 
become 20\% worst of that obtained from the fit with the scaled PSF. 
 Since we have no knowledge of the underlying host galaxy we performed this evaluation of 
the brightness limit of the host galaxy using different models. 
We assumed two type of morphology : exponential disk and de Vaucoulers laws. 
Then we assume two values for the half light radius: 5 and 10 kpc that are representative values of the 
resolved objects in our sample 
The input half light radius in arcsec was derived assuming the redshift of the QSO.
We took as upper limit to the luminosity of the host galaxy the maximum value derived from the various fit 
using different models and half light radius.

The final classification of the targets was based on the comparison of $\chi^2$ for the two  
fit (only psf and psf + galaxy) and further visual inspection of the fit. From this procedure we classified 
all objects as resolved, unresolved and marginally 
resolved for intermediate fit ( see Figure \ref{fig:chi2ratio} and Table \ref{tab:results}).

\section{Host galaxy properties}
\label{sect:hostgal}
\begin{figure}
\centering
\includegraphics[bb=70 170 520 670,width=0.8\columnwidth]{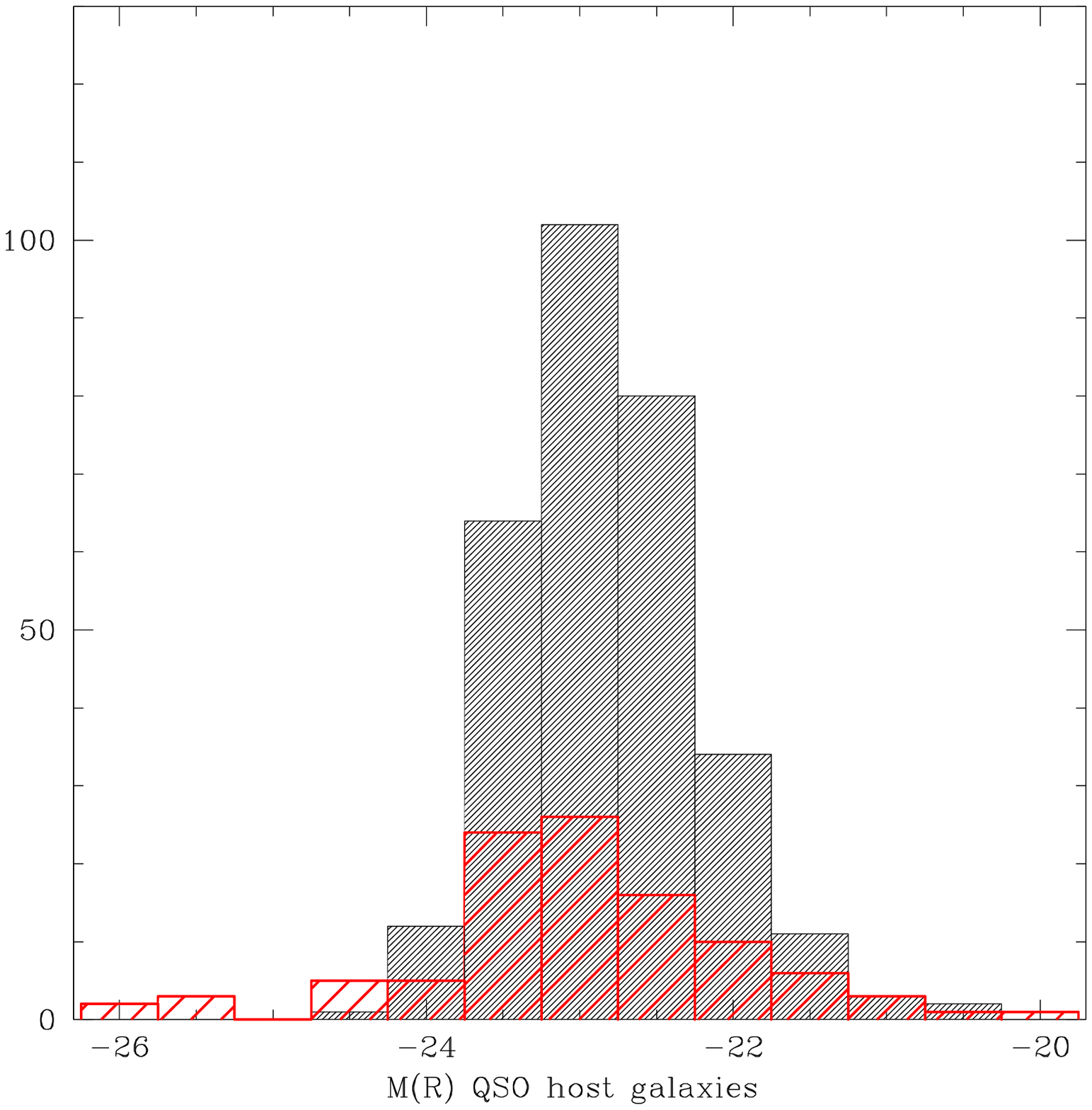}
\caption{Absolute magnitude distribution of resolved  QSO  host galaxies in the rest frame R band. 
For comparison the distribution for a compilation of low redshift QSO 
imaged by HST is plotted (dashed red region) \citep{decarli2010b}. } 
\label{fig:mhabsR}
\end{figure}

In Figure \ref{fig:mhabsR} we show the distribution of the absolute magnitude of host galaxies  for all (309)
resolved  quasars. These values were corrected for the galactic extinction (based on the values of SDSS) and k-corrected to the R band rest
frame. 

\begin{figure}
\centering
\includegraphics[bb = 50 170 520 510, width=0.9\columnwidth]{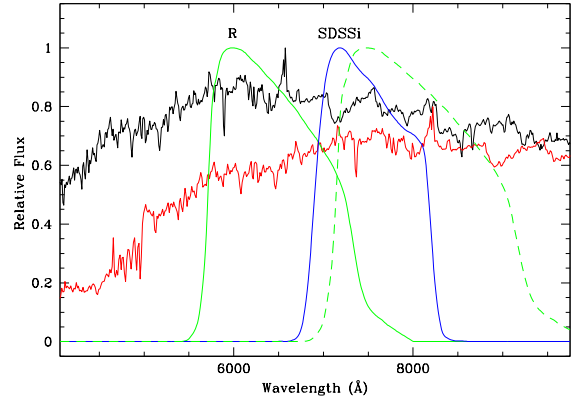}
\caption{  Example of k-correction and filter transformation for an object at 
z = 0.25 assuming a SED of an elliptical galaxy. The template for the elliptical galaxy is shown at rest frame (black line) and at z=0.25 
(red line). The transmission for passbands R (solid green line) and i (blue line) are compared with the redshifted (z=0.25) 
R passband (dashed green line). 
}
\label{fig:kcorr}
\end{figure}

To perform the color and
$k$-correction transformations, we assumed an elliptical and early type spiral galaxy spectral energy distribution (SED) 
template (\cite{mannucci01}) for the host galaxy. Because of the small difference of the templates in the 
observed spectral region, the k-correction adopting the two SEDs differ by few percent, thus we used 
for all objects the one for elliptical SED. 
For the nucleus we used a composite quasar spectrum (Francis et al. 1991).
All k-corrections were performed adopting these templates for the SED and convolving them with the $i$ 
and R filter responses.
An example of the procedure is illustrated in Figure \ref{fig:kcorr}

The average absolute magnitude is $<M(R)>$ = -22.83 $\pm$ 0.6 (median -22.86 $\pm$ 0.36) . For
comparison the distribution of absolute magnitude for a smaller ($\sim$ 100 ) sample of QSO observed by HST (see
compilation by \cite{decarli2010b}) in the similar redshift range is $<M(R)>$ = -23.00 $\pm$ 1.05. The two data
sets are in excellent agreement in spite of the differences of observation technique.
Five quasars in our sample were observed with Hubble Space Telescope and WFPC2 in filter F606W 
\cite{cales11} and it is possible to compare our analysis with the results from HST images.
The comparison of the magnitudes of these host galaxies (assuming a color correction V-R = 0.8) 
is very good  (  $< \Delta m>$ = 0.1 $\pm$ 0.24 )

% --- FIG 8 ----- QSO host galaxy distribution 
%\clearpage

\begin{figure*}
\centering
\includegraphics[ width=1.95\columnwidth]{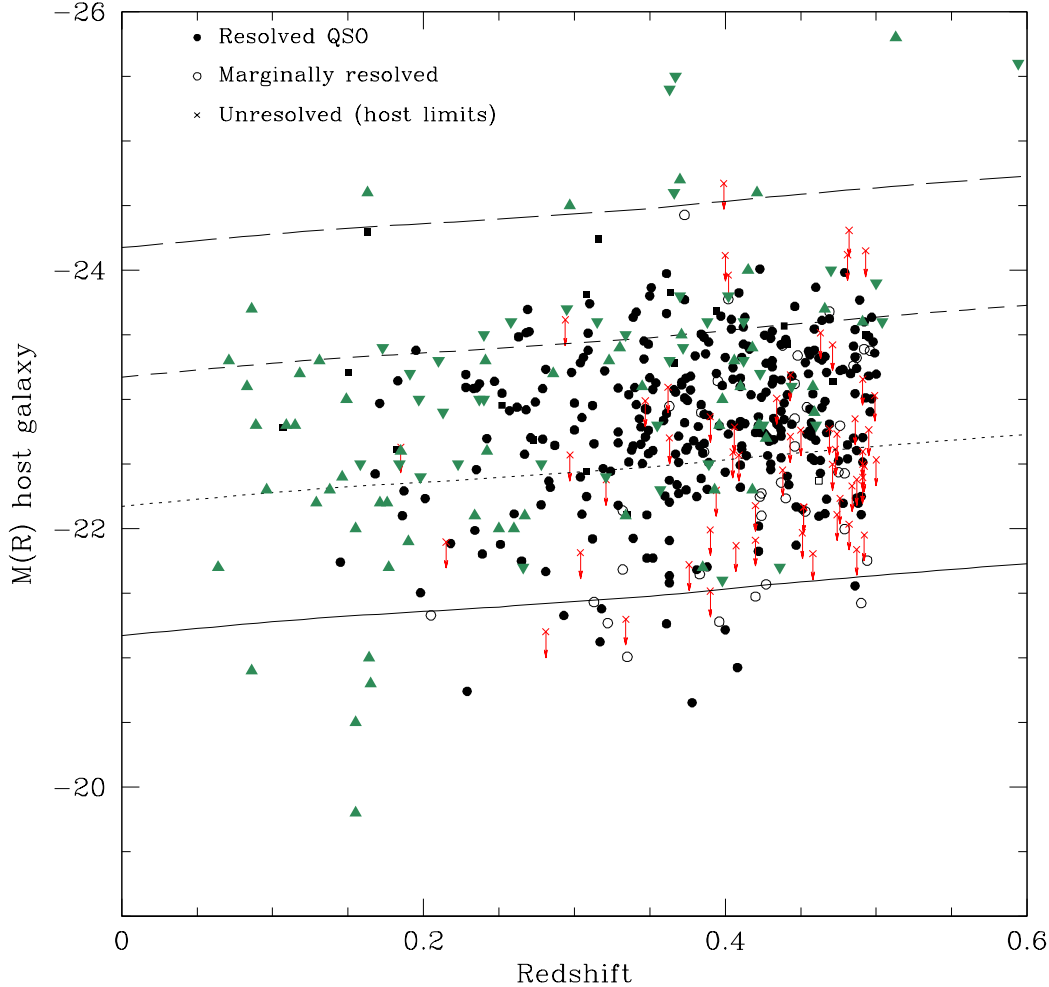}
\caption{The absolute magnitude of QSO (RQQ circles; RLQ squares) host galaxies versus redshift. 
Resolved quasars (filled points), marginally resolved (open points) and 
luminosity lower limits (red crosses with arrows). For comparison we include 
a compilation of $\sim$ 100 QSO host galaxies from HST observations  \citep{decarli2010b}
(filled green triangles: inverted triangles for radio loud objects ) } 
\label{fig:mhz}
\end{figure*}

The distribution of host galaxies in the redshift-luminosity plane (see Fig \ref{fig:mhz}) confirms previous
claims that they are encompassed between M* ( M*(R) = --21.2 ;  \cite{nakamura03}  )  and M* - 3 with more frequent distribution in the range  M*-1 and M*-2  \citep{kukula01,dunlop03,falomo04,peng06,kotilainen07,decarli2010b}.

There is a small, but significant, increasing of the host  luminosity with the redshift (from M(R) $\sim$ -22.5
at z $\sim$0.2 to M(R) $\sim$ -23.1 at z $\sim$ 0.5 ) that is consistent with passive evolution of the
underlying stellar population. A similar trend was also reported over a wider redshift range by \citet{
kotilainen09}.

\begin{figure}
\centering
\includegraphics[bb=70 150 530 660,width=0.95\columnwidth]{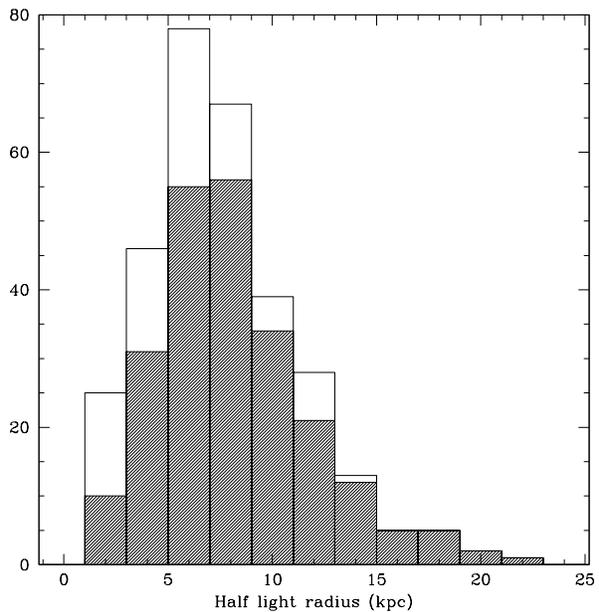}
\caption{Distribution of half light radius of QSO host galaxies as 
derived from the fit with a Sersic law plus nuclear component (open histogram).
In the shaded area the distribution for well resolved QSO is shown (see text).
Only objects classified as resolved are included.
} 
\label{fig:hlr}
\end{figure}

While the total flux from the host galaxy is relatively well determined from the fit of resolved objects the
evaluation of the half light radius is not well constrained. This is due to some degeneracy between the effective 
surface
brightness and effective radius that can be combined to produce the same total flux. In Figure \ref{fig:hlr} we
show the distribution of the effective radius (Re) as derived from the fit to the objects with a galaxy (modelled
by a Sersic law)  plus the nuclear component (modelled by PSF). The average effective radius is $< Re >$ = 7.7
$\pm$ 3.6 kpc. If we include only the targets with $good$ fit ($\chi^2$ /  $\chi^2$(PSF) $<$ 0.5) and
uncertainty of Re $<$ 30\% the average effective radius is slightly larger : $< Re >$ = 8.2 $\pm$ 3.7 kpc. For
the small sample of radio loud quasars (17 objects) the average effective radius ($< Re >$ = 8.8 $\pm$ 5.3 kpc )
is indistinguishable from that of the whole sample. In a hierarchical scenario of galaxy formation where one
would expect that the size of galaxies evolve with the cosmic time as was reported in a number of detailed
observations of galaxies at different redshift (e.g. \citet{bouwens02,bouwens04,trujillo06,ono12}) the size of
the galaxies (as derived from the half light radius) ranges from compact (few kpc) objects up to extended
galaxies (10-15 kpc); in the observed redshift range we do not find any significant trend  of change of the
galaxy size with z. In some  cases we found a significantly larger than average galaxy  
radius (see Fig. \ref{fig:hlr}) that is likely due to the presence of an extended halo. 
These, and other peculiarities will be investigated in \cite{bettoni14}. 

\begin{figure}
\centering
\includegraphics[bb=25 150 530 660, width=0.95\columnwidth]{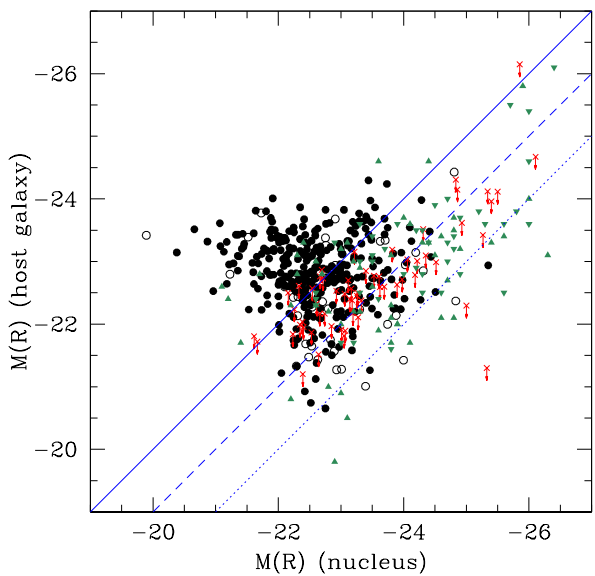}
\caption{ The comparison between the nuclear and host galaxies luminosities in the R band.
Resolved quasars (filled circles), marginally resolved (open circles) and 
luminosity upper limits (red crosses with arrows). For comparison we include 
a compilation of $\sim$ 100 QSO host galaxies from HST observations  \citep{decarli2010b}
(filled green triangles: inverted triangles for radio loud objects )
Diagonal blue lines represent the loci of fixed ratio between the nucleus and host galaxy luminosity (in the R band) 
at constant levels of 1 (solid line), 2.5 (dashed line), 6.25 (dotted line).
} 
\label{fig:mhn}
\end{figure}

An interesting controversial issue in the study of QSO and galaxies is  the relationship between the  nucleus and host
galaxy luminosity. Assuming that quasars emit in a relatively narrow range of Eddington ratio and that the  BH mass is
correlated with the mass of the galaxy one would expect to find a correlation between nucleus and host galaxy luminosity.
In figure \ref{fig:mhn} we show the comparison between the absolute magnitude of the nucleus and that of the host galaxy as
derived from our image analysis for resolved and marginally resolved objects. Both luminosities were k-corrected and  refer
to rest frame R band. With our QSO sample we can explore a range of nuclear luminosity between M(R) $\sim$ -20 and M(R)
$\sim$ -24  (average $<$M(R)$_{nuc}$ $>$ = -22.58 $\pm$ 0.80). In this luminosity range there is not  a significant
correlation between the two quantities (see Fig. \ref{fig:mhn}). The same result is derived if we include the compilation
of low redshift quasars observed by HST \citep{decarli2010b} that extends to higher QSO luminosity. The only exceptions to
this behaviour appears to be for few high luminosity (M(R) $\sim$ --26) quasars that are hosted in very high luminosity
galaxies. A similar behaviour was also noted by \cite{mcleod00} from the comparison of a collection of Seyfert galaxies and
low z QSO data and interpreted as a luminosity/host-mass limit.  If applied to our sample this suggests that a limit is
reached when the nucleus emits (in the R band) a power corresponding to a factor 3-5 higher than the luminosity of the
whole host galaxy. The same behaviour was also observed, albeit in a smaller sample, for high redshift quasars \citep{
kotilainen09} and confirms that an intrinsic range of accretion together with different mechanisms for low power emission
may concur to destroy the correlation. Moreover it is worth to note that if the BH mass is related only with the bulge mass
/ luminosity then one would expect additional disruption of the above correlation between BH and galaxy masses (see also discussion in the next
section).

\subsection{Host galaxy morphology}

A long debated question concerning the properties of the galaxies hosting quasar is 
its morphology (see e.g. \cite{bohm13} and references therein).  
Do quasars inhabit both disc and bulge dominated galaxies ? 
This question was debated for long time since the poor spatial resolution of 
the observations combined with the bright nuclei hindered the clear nature of the QSO hosts.
The original idea  that considered radio loud QSO being hosted by ellipticals 
while radio quiet quasars hosted in spiral galaxies is clearly not consistent with the observations 
that show a more complex scenario.

In the era of HST images it was clear that at low redshift QSO are found in both types of galaxies 
spiral and ellipticals and also in complex morphology and interacting galaxies 
\citep{bahcall97, kukula01, ridgway01}. 
It was also suggested that there may be a relationship between QSO luminosity and host galaxy morphology
such that all the radio-loud quasars, and all the radio-quiet quasars  with nuclear luminosities M$_V <$
-24, are massive bulge-dominated galaxies, \citep{floyd04}.

%\bigskip
%--- FIG 7 ---
%--- FIG --- HLR -------------------------------------------------

% ---  OUR RESULTS about morphology of host galaxies

From the analysis of our large dataset we found that 309 out of 416 targets are well 
resolved (see above discussion), however, to be able to constrain the morphology 
of the host galaxy it is needed that the flux from the surrounding nebulosity is well detected up to 
large radii (faint surface brightness) where the two models ($r^{1/4}$ and exponential disk) differ 
significantly. 
In order to classify the morphology of the host galaxies we can use the values of 
Sersic index obtained from the best fit. 
In addition  we also performed a fit of all objects assuming  the host galaxy is 
a pure elliptical or a pure disk and then compared the $\chi^2$ of the two fit. 
This kind of analysis can yield only a preliminary indication of the morphology of the host galaxies 
since in general both spheroidal and disk components may be present. 
In order to better characterise the host galaxies we performed a detailed 
visual inspection of all resolved  targets using the whole information available: 
images, contour plots, fit of the brightness profile, ellipticity. 

The morphological classification of galaxies is an important indicator of many physical
processes in galaxies and the Galaxy Zoo project is a clear example \cite{linlott11} 
although it is somewhat a subjective process. For instance note 
 that  the various tools for  automatic classification  or the Galaxy zoo web based one can 
 only give a rough estimate of the morphology (see e.g. \cite{nair10}  for full discussion ). 
For this reason we used the recipe of \cite{nair10} to classify our
host galaxies in the same classification scheme of the Carnegie Atlas of Galaxies \citep{sandage94} 
and of RC3 (\cite{rc3}). Due to the presence of the central
nuclear source we restricted our T types to five main classes T=-5,-4 for all early type
galaxies, T=-1 for E/S0 T=0 for S0 galaxies and T=1,2 for late type galaxies. Our
classifications take into account both the visual inspection of the i band image and the
luminosity profile.

From this analysis we find  that the morphology of the host galaxies is rather complex with both disk and spheroidal
components often present in these galaxies. Of the 314 resolved targets about 113 objects (37\%) are dominated by the bulge
component, 129 objects (42\%) have a conspicuous disk structure, 64 objects (21\%) exhibit a mixed (bulge plus disk)
features. For $\sim$100 objects (32\%) a number of complex features (lens, tidal distortion, bars, close companions) are also
present. Detailed analysis of these morphological structures will be presented in another paper of this series \citep{
bettoni14}. This morphological classification of the host galaxies is summarised in  Table \ref{tab:abs}.

\section{Black hole mass and host galaxy relationship}

Massive black holes (BHs) are ubiquitously found in the centre of massive galaxies and their
masses  show correlations with large-scale properties of the host galaxies, namely, the
stellar velocity dispersion, the luminosity, and the mass of the spheroidal component \citep{ferrarese06,marconi03, 
bernardi07,letawe10}. These
relations have been interpreted as the outcome of a joint evolution between BHs and their
host galaxies and are therefore potentially of great importance for the understanding of the
processes that link nuclear activity to galaxy formation and evolution 
\citep{jahnke09,decarli2010b,merloni10,targett12,cisternas11}.

Our large and homogeneous dataset allows us to investigate this relationship for low redshift quasars. For the BH mass we
adopted the measurements obtained by \cite{shen11} who estimate the virial BH mass using the FWHM of $H_{\beta}$ and
continuum luminosity (\cite{Vestergaard2006}) for all the QSO in SDSS DR7 with $z<0.7$ (\cite{Vestergaard2006}). All the
spectra in our sample were  visually inspected and 31 objects ($\sim$7\%) have been removed from the  sample because of
very low S/N ratio of the spectra. For three cases (i.e. objects 311, 349 and 365) we have done a new measurement of the
BH mass.  Since these BH masses are derived from single-epoch virial mass estimates and assume an indirect measurement
of the size of the BLR from its  relationship with the  continuum luminosity the individual values may have large
uncertainty. For our QSO sample the quoted errors of BH mass by \cite{shen11} taking into accounts various effects (see
\cite{shen11} for details) range on average from 0.1 dex up to about 0.4 dex (mean error 0.17 dex) 
with even larger errors in few  cases. Note
that this uncertainty includes neither the statistical uncertainty ($\geq$0.3-0.4dex) from virial mass calibrations, nor
the systematic uncertainties with these virial BH masses. 
 
\begin{figure}
\centering
\includegraphics[bb=70 150 530 660, width=0.95\columnwidth]{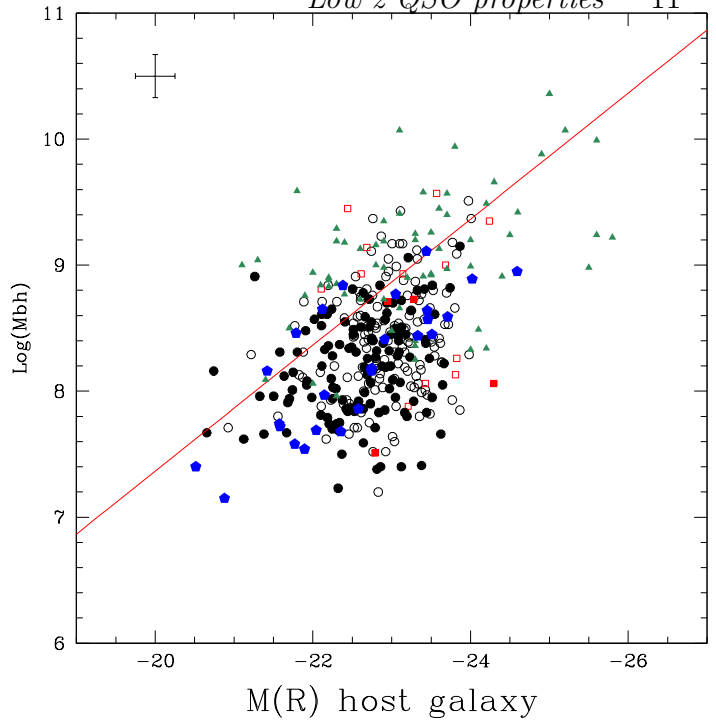}
\caption{Absolute magnitude of QSO host galaxies versus BH mass for resolved 
quasars. The reference (red) line is the Bettoni et al. (2003) relation for local (inactive) galaxies for which 
black hole mass was measured. Open points are QSO with poor spectra and more uncertain BH masses.
Red points are radio loud quasars. The mean uncertainty on BH masses 
is given by the error bar on the top left of the figure (see text for details).
For comparison we include the compilation of QSO (green triangles) with z $<$ 1 
from Decarli et al. 2010a. The  blue pentagons represent the sample of  quasars 
observed with HST \citep{bentz09a,bentz09b} and for which the BH mass was derived from 
reverberation mapping technique \citep{cales11}.}
\label{fig:mbhmh1}
\end{figure}

\begin{figure}
\centering
\includegraphics[bb=40 150 520 650, width=0.95\columnwidth]{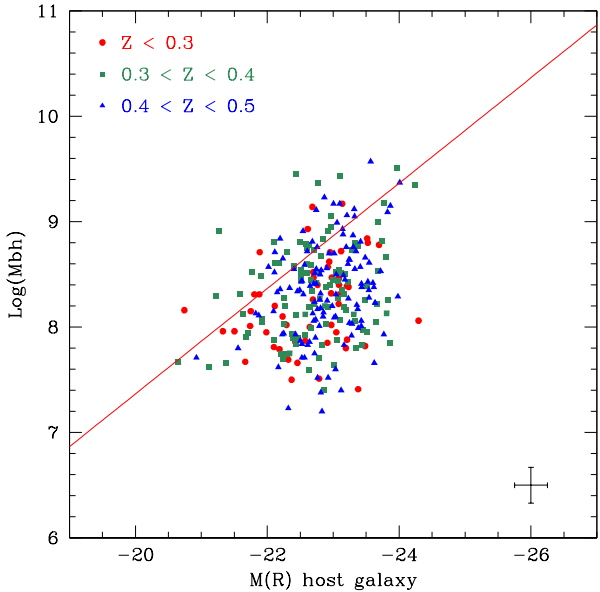}
\caption{Absolute magnitude of QSO host galaxies versus BH mass for resolved 
quasars in different redshift bins. No significant difference is found among the various redshift ranges.
The mean uncertainty on BH masses 
is given by the error bar on the bottom right  of the figure (see text for details).}
\label{fig:mbhmhz}
\end{figure}

\begin{figure}
\centering
\includegraphics[bb=70 150 530 660,width=1.0\columnwidth]{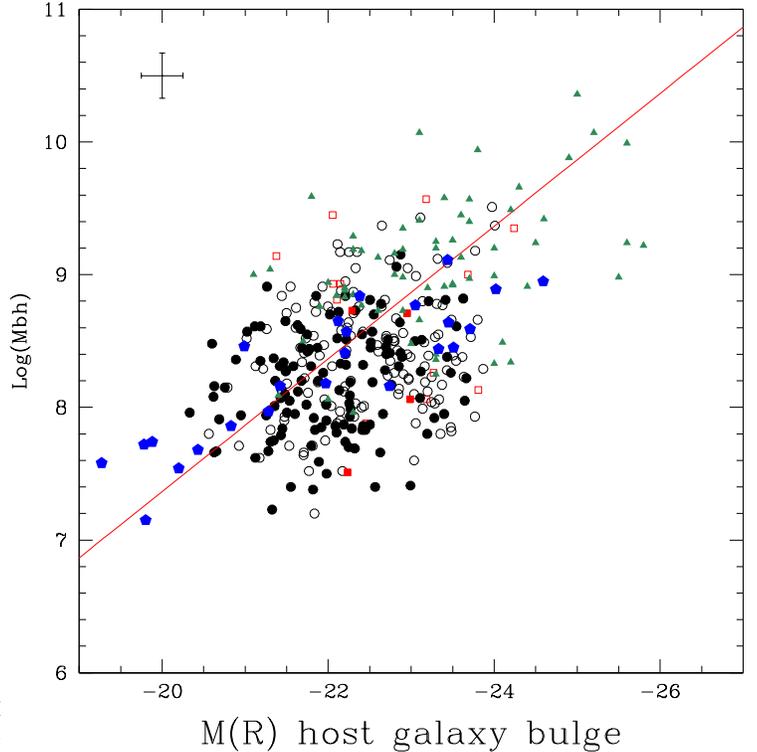}
\caption{ Same as Figure \ref{fig:mbhmh1} but for the estimated host galaxy bulge component. }
\label{fig:mbhmh2}
\end{figure}

In Figure \ref{fig:mbhmh1} we report the relationship between the black hole mass (M$_{BH}$) and the absolute
magnitude of the host galaxy in the R band for all resolved quasars that have good S/N spectra (see above). The
absolute magnitude M(R) of the host galaxies are in the range -22 to -24 and BH masses between 10$^7$ to 
10$^9$ M$_o$. 
We search for possible evolution with the redshift of the  M(R) - M(BH) relation and report in Figure \ref{fig:mbhmhz}
the comparison of the relation for  different redshift intervals. From our dataset we do not find any significant evolution 
of the M(BH) - M(R) relation from z $\sim$ 0.2 to z $\sim$ 0.5.
On average the BH masses are found systematically lower than the 
value expected for the host galaxy luminosity with respect to the M(R) - M(BH) relation  established for 
local (inactive) galaxies. 
In addition there is a large scatter of BH masses at the same galaxy luminosity.
For the galaxies of absolute magnitude -22 $<$ M(R) $<$ -24  the BH mass is 
spread over about 2 order of magnitudes. 
We argue that  this is an indication that the BH mass is not well correlated with the total mass of the 
galaxy. If the BH mass is linked only with the spheroidal component  then the correlation would be 
significantly improved.
Indeed we have many quasars that are hosted in galaxies with a significant disk component therefore 
their bulge (or spheroid component) represent only a fraction of the total luminosity 
of the galaxy. In Figure \ref{fig:mbhmh2}  we show the M(BH) -- M(R)  (estimated bulge) 
relation taking into account the effects of disk/bulge components. 

 For each resolved quasar we have classified the morphological type of the host galaxy 
as described in Section  4. 
We have then evaluated the fraction of the bulge to total galaxy luminosity in a range between 1 to 0.3 following the 
above morphological classification.  In Figure \ref{fig:mbhmh2} we show  the relationship between M(BH) and 
the estimated bulge luminosity that exhibits a significant correlation. 
This indicates that the BH mass is linked only with 
the bulges mass/luminosity and not (or only modestly) with the total mass of the galaxy. 
 This result is also supported by the comparison of the M$_{BH}$ host galaxy relation for 
25 low redshift (z $<$ 0.2) quasars  \cite{bentz09a, bentz09b} for which the 
BH mass was derived from reverberation mapping technique and the host galaxy properties were 
obtained from ACS HST images. For about half of these objects the quasar host galaxy exhibits a 
significant disc component   and a bulge to total galaxy ratio was derived \cite{bentz09b}.
For the rest of objects a pure bulge (elliptical) component  was derived.
It turns out that these data well overlap with our relationships for M(BH) - host galaxy (see Figure \ref{fig:mbhmh1} ) 
and M(BH) - bulge (see Figure \ref{fig:mbhmh2} ) 
It is worth to note that at the lowest BH masses we find QSO with relatively luminous host galaxies 
(see Figure \ref{fig:mbhmh2} ) while those observed by Bentz et al., at similar BH masses, are 
significantly less luminous and lie close to 
the local M(R) - M(BH) relation. Since our QSO with small BH masses are well resolved it is 
unlikely that their host luminosity be overestimated. 
On the other hand note that the BH masses in Bentz et al. are measured by reverberation mapping 
that resolves the influence of the black hole in the time domain through spectroscopic
monitoring of the continuum flux variability and the delayed response,  in the broad emission-line flux.
We thus argue that some bias could be in place using the virial method to derive M(BH) for low luminosity 
(likely lowest BH masses) QSO.

% Figure to show there is no evolution of MBH - MH relation

\section{Summary and Conclusions}

We have investigated  the properties of the host galaxies from a large ($\sim$ 400 objects) and homogeneous dataset 
of low redshift (z $<$ 0.5) quasars using th SDSS images in the Stripe82 region that are significantly deeper that 
standard SDSS data. The 2D analysis of the images allowed us to well resolve the 
quasar host for  3/4 of the objects in the sample, marginally resolve other 40 quasars 
and derive limits for the galaxy luminosity for the unresolved targets (60 objects).

The following  properties of quasar hosts are derived:

\begin{enumerate}
\item the luminosity of the host galaxies of low z quasars span a range from M(R) $\sim$ -21.5 to M(R) $\sim$ -24.0; the
bulk of the host galaxies are located in the region corresponding to M*-1 and M*-2; there is a mild increase of the host
luminosity with the redshift that is consistent with the passive evolution of the underlying stellar population

\item the morphology of the host galaxies turned out to be rather complex with both bulge and disc dominated 
galaxies; about one third of the objects in our sample show features characteristics of bulge and disc components

\item irrespective of the host morphology the size of the galaxies (as derived from the half light radius) ranges from
compact (few kpc) objects up to extended galaxies (10-15 kpc); in the observed redshift range we do not find any
significant trend  of change of the galaxy size with z

\item  the nuclear and host galaxy luminosities are not 
correlated suggesting that accretion rate, BH mass, and galaxy masses and morphology 
combine together to smear significantly the correlation between BH and host masses  

\item the BH mass of quasars estimated from the QSO continuum luminosity and the width of the broad emission lines 
is poorly correlated with the total luminosity/mass of the whole host galaxy; on the contrary when the fraction of 
bulge to disc component is considered we find a significant correlation between the BH mass and 
the bulge luminosity of the host.

\end{enumerate}

Another important source of information to characterise the properties of low redshift QSO come from the analysis 
of their galaxy environments as compared with those of similar galaxies with no active nuclei. 
These aspects will be pursued in forthcoming papers of this series \citep{bettoni14,karhunen13}.

% ----------------------------------------------------------------
\section{Appendix}
\label{app1}
In order to test the reliability of the image decomposition we have performed a number of 
mock simulations of quasars and then analysed them with the same method used for the Stripe82 images.
To perform the simulation of the quasars we used the Advanced Exposure Time Calculator tool \footnote{AETC available at 
http://aetc.oapd.inaf.it/ }. The parameters of the Sloan telescope and the global 
efficiency of the instrument were adopted from SDSS web site. We used read out noise of 9.5e, gain of 3.8 \citep{gunn98}, exposure time of 1 hour (subdivided in 60 exposures of 1 minute), 
sky brightness as average value of Kitt Peak. Statistical noise was added to the mock objects and background. 
The final images are background subtracted as are Stripe82 images.

Three set of simulations were performed assuming different galaxy model with Sersic index n=1, 2.5 and 4 to represent 
disk, intermediate and elliptical host galaxies , respectively. For each dataset we used 
a number of PSF extracted 
from the PSF of our images and  corresponding to seeing in the range 1.0 -- 1.4 arcsec (FWHM).
Then we construct the QSO images as the superposition of a nucleus and a galaxy with a range of 
values that map the observed values. We explored a range of nucleus/host galaxy flux ratio from 0.1 to 2.
Moreover for each galaxy model we assume effective radius of 1.0, 1.5, 2.0 and 2.5 arcsec 
(again similar to the observed parameters).

\begin{figure*}
\centering
\includegraphics[bb= 40 10 810 750 ,width=1.9\columnwidth]{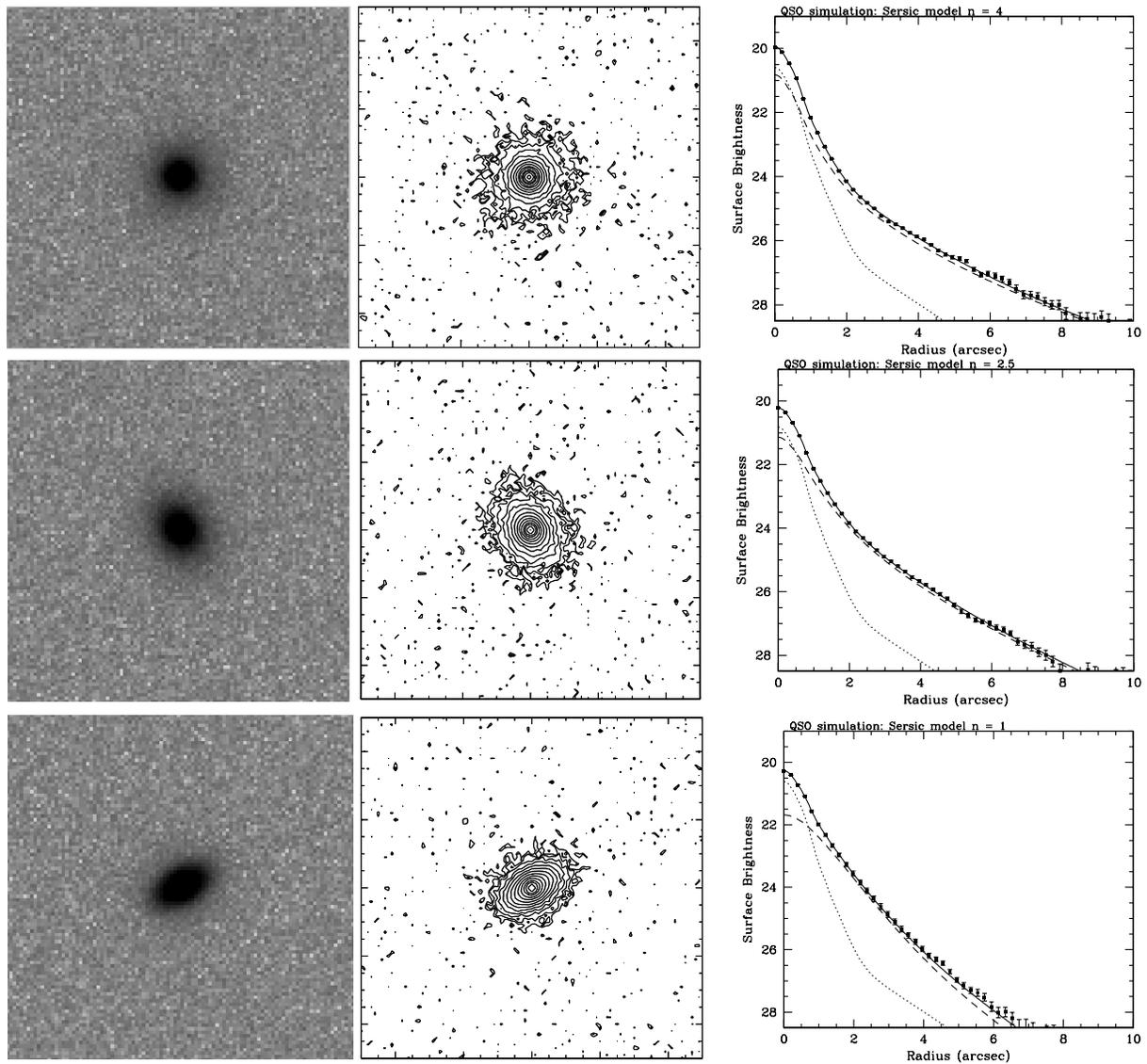}
\caption{Example of mock simulations of QSO (nucleus + host galaxy) with a Stripe82 PSF 
corresponding to 1.2 arcsec and  nucleus/galaxy flux ratio of 0.5, and effective radius of 1.5 arcsec.
Left panels show the simulated images; central panels the contour plot and the right panels 
the best fit of the QSO image.
Top: Sersic index n = 4, ellipticity 0; Middle: Sersic index n = 2.5, ellipticity 0.3; 
Bottom: Sersic index n = 1, ellipticity 0.5.
}
\label{fig:mock_comb}
\end{figure*}

\begin{figure}
\centering
\includegraphics[bb=30 290 290  700,width=0.49\columnwidth]{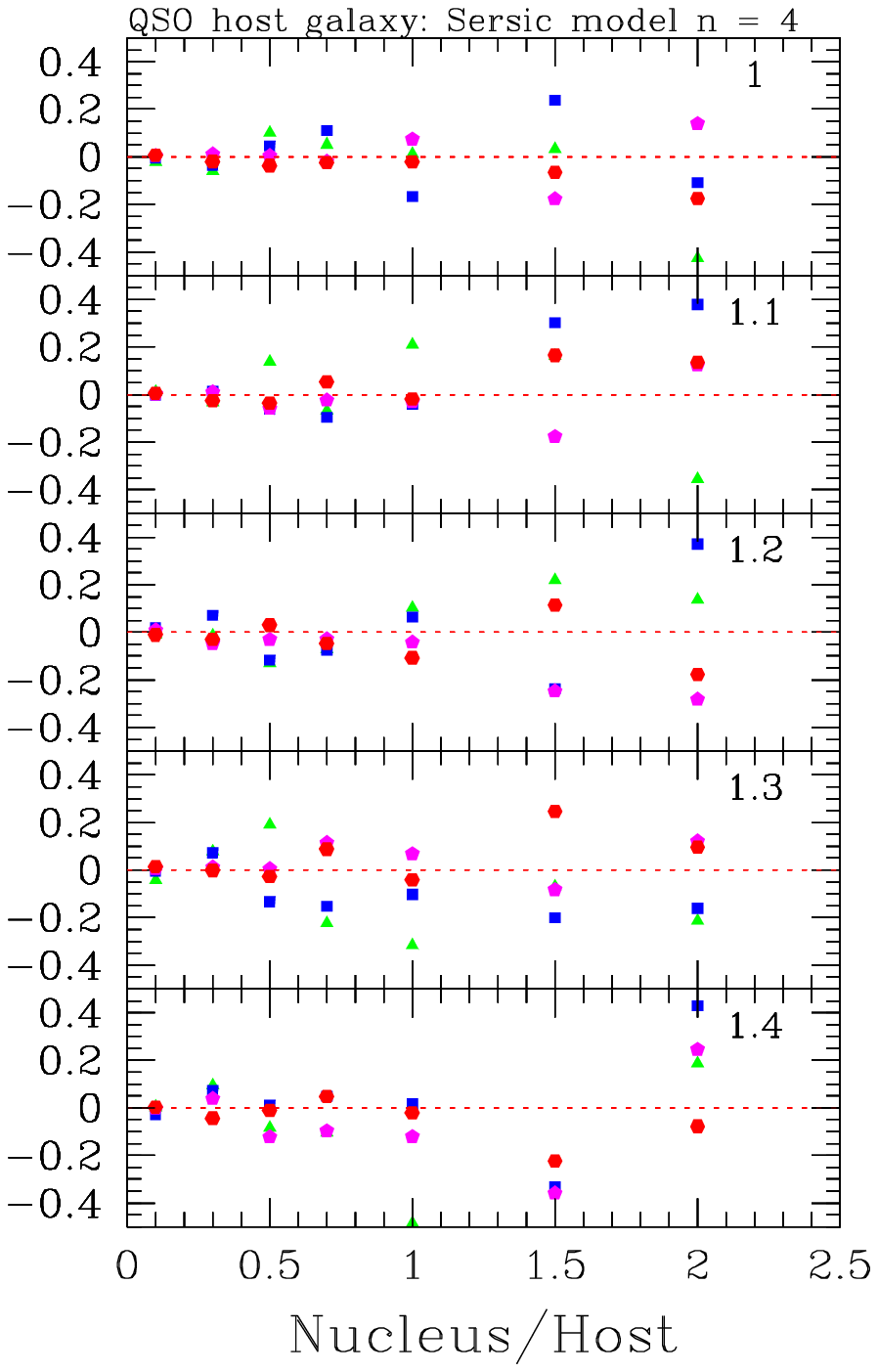}
\includegraphics[bb=30 290 290  700,width=0.49\columnwidth]{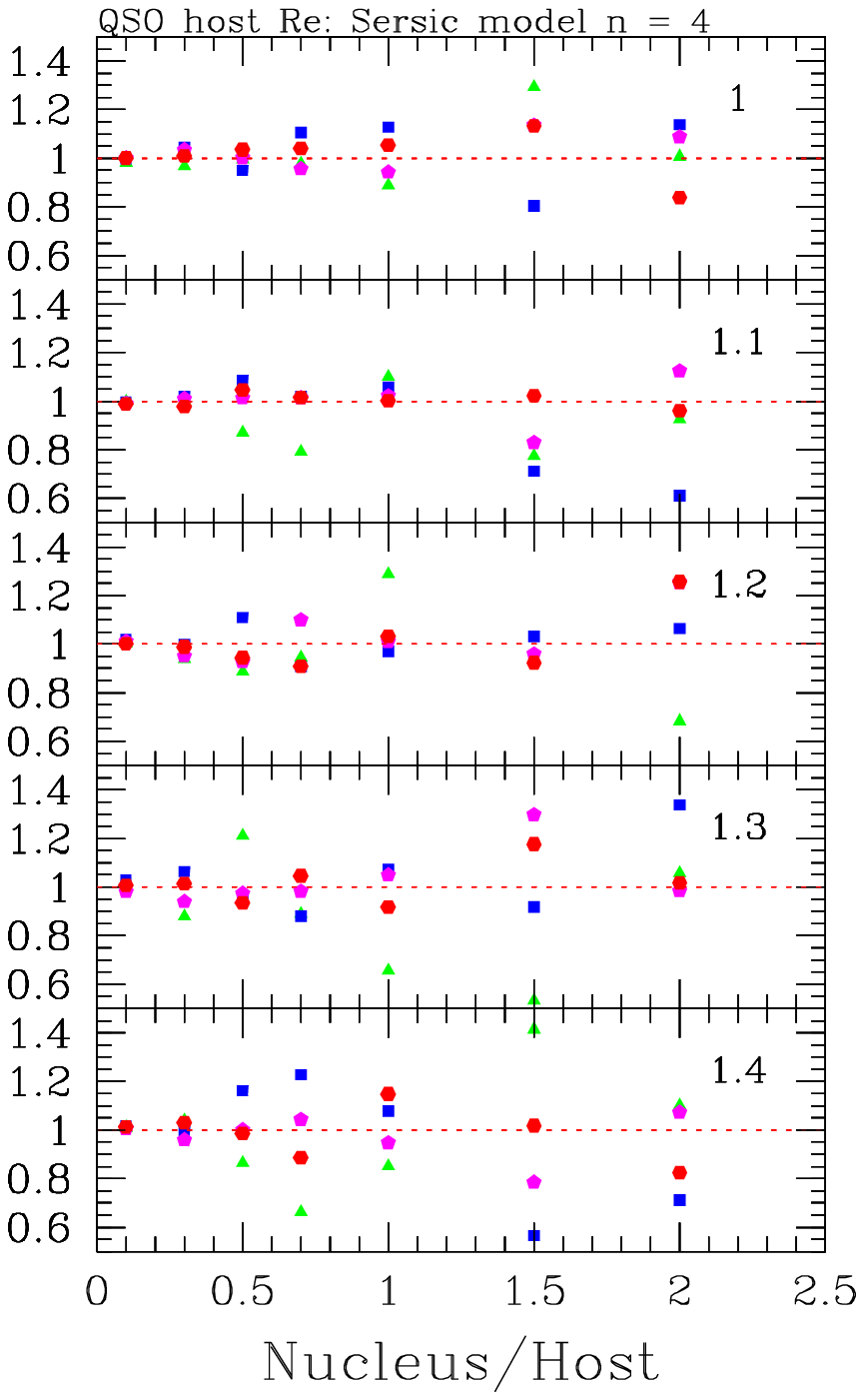}
\caption{ Mock simulation of QSO (nucleus + host galaxy with Sersic index n = 4). 
Left panels: Difference of the host galaxy magnitude 
between measured and input galaxy as a function of the nucleus/galaxy flux ratio. 
The panels represent simulations assuming  
different PSF of seeing between 1.0 and 1.4 (top right of each sub panel). 
The host galaxy is simulated with effective radii of 1.0 arcsec (green triangles), 1.5 arcsec (blue squares), 2.0 arcsec 
(magenta pentagons) and 2.5 arcsec (red hexagons). Right panel show the ratio between 
the measured effective radius and the true effective radius as a function of nucleus/galaxy flux ratio for the same combination of 
PSFs and effective radii.
}
\label{fig:mock_n4}
\end{figure}

\begin{figure}
\centering
\includegraphics[bb=30 290 290  700,width=0.49\columnwidth]{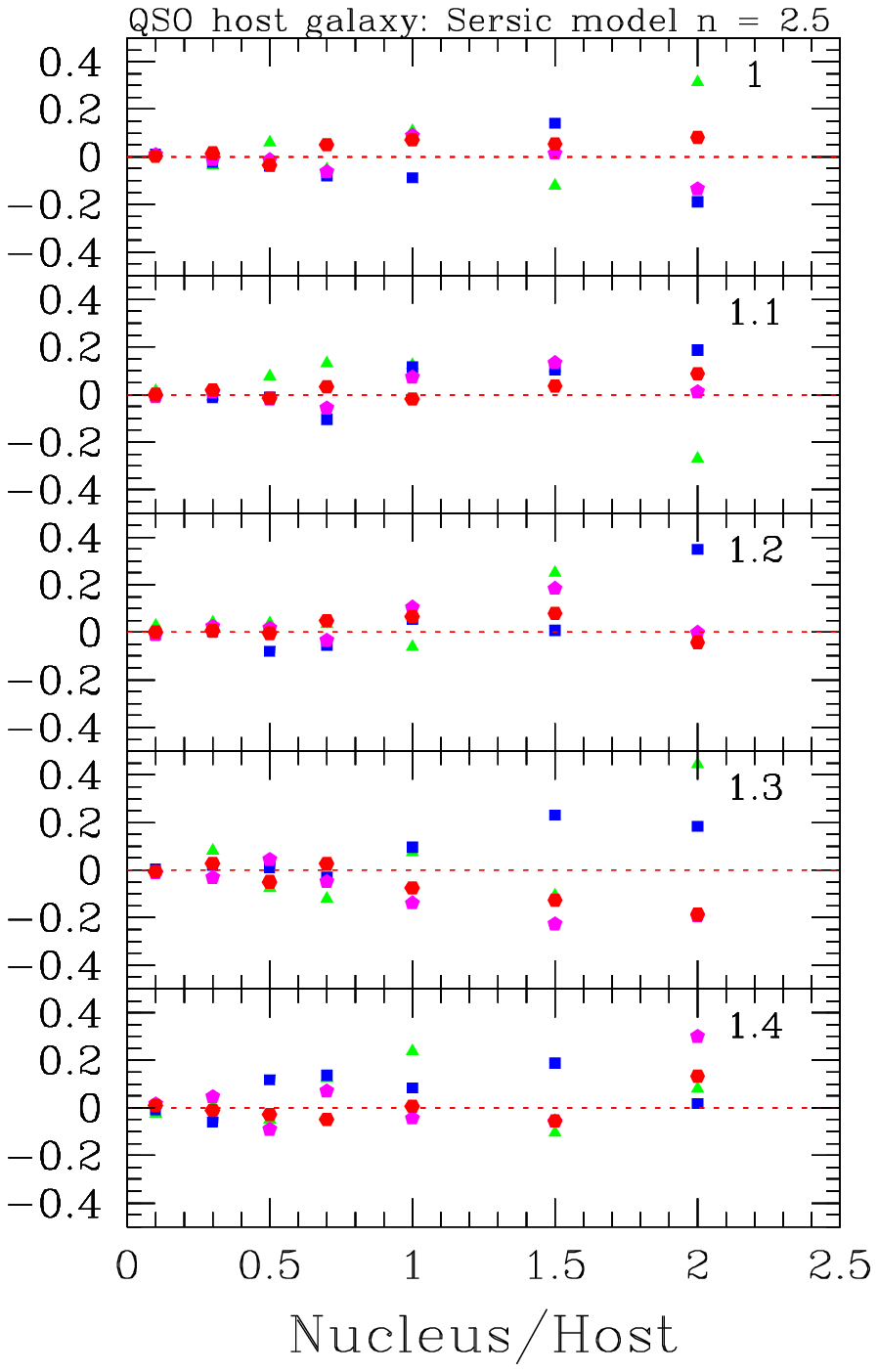}
\includegraphics[bb=30 290 290  700,width=0.49\columnwidth]{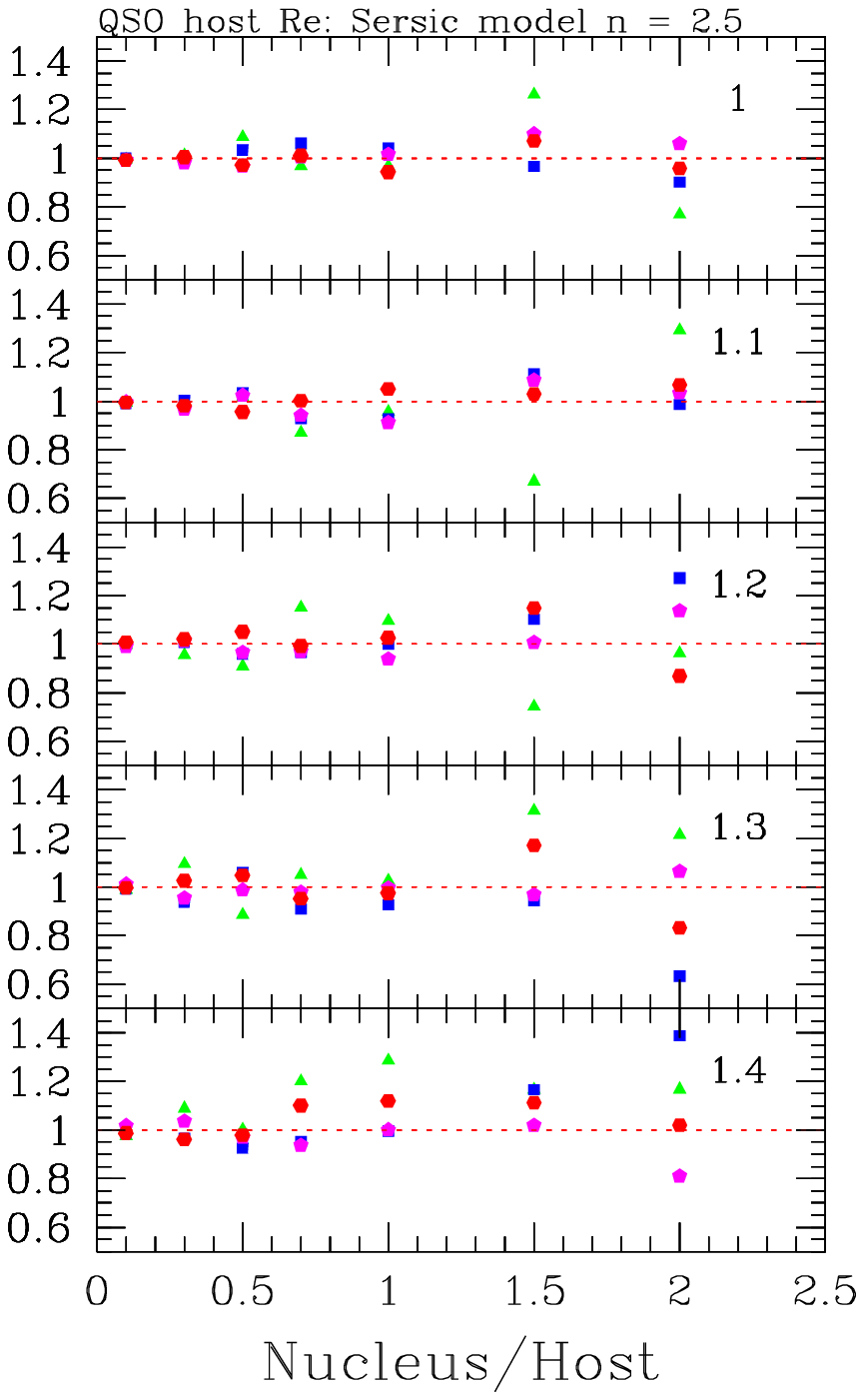}
\caption{ Same as Figure \ref{fig:mock_n4} but for host galaxy with Sersic index n = 2.5. }
\label{fig:mock_n25}
\end{figure}

\begin{figure}
\centering
\includegraphics[bb=30 290 290  700,width=0.49\columnwidth]{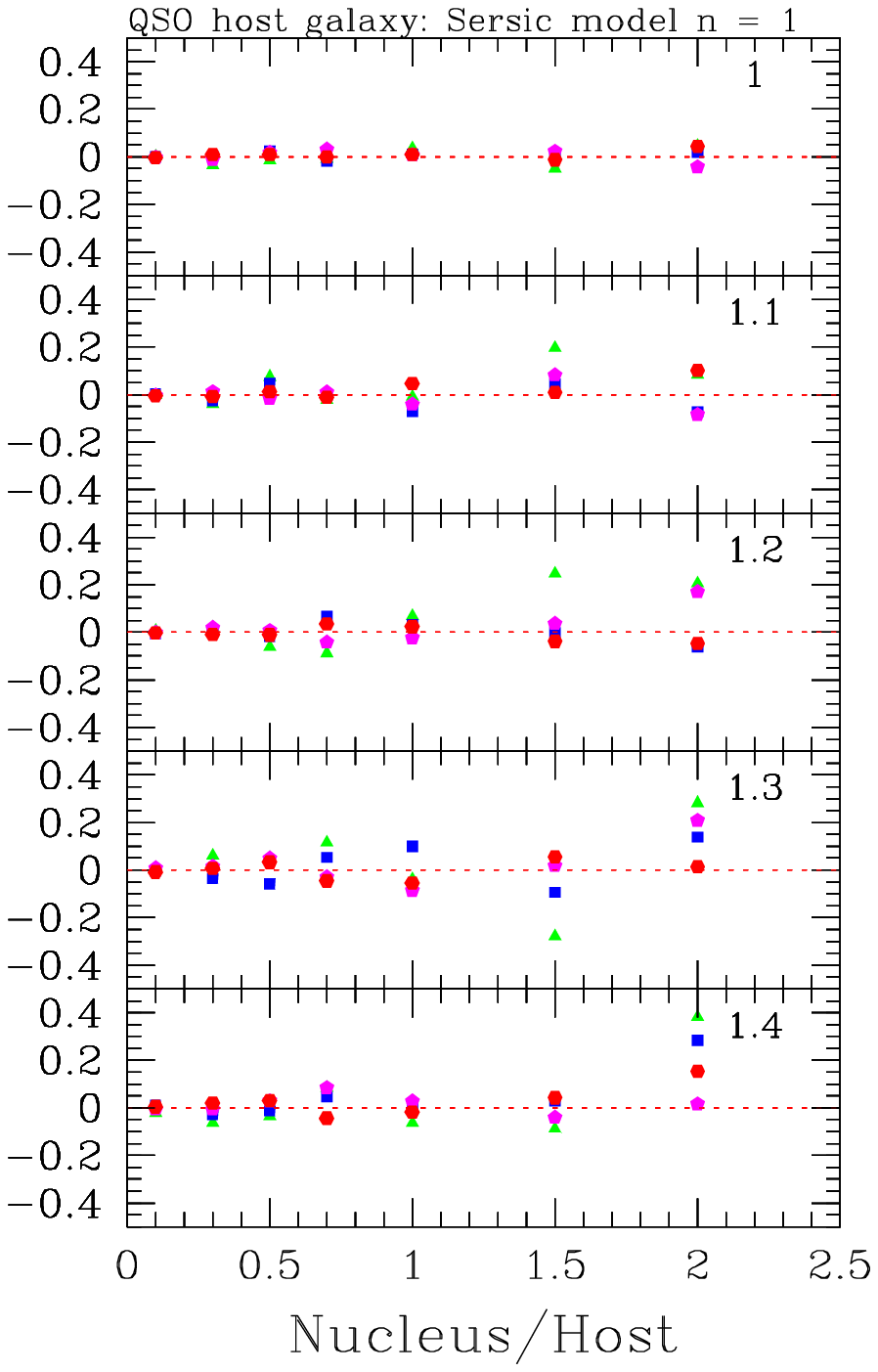}
\includegraphics[bb=30 290 290  700,width=0.49\columnwidth]{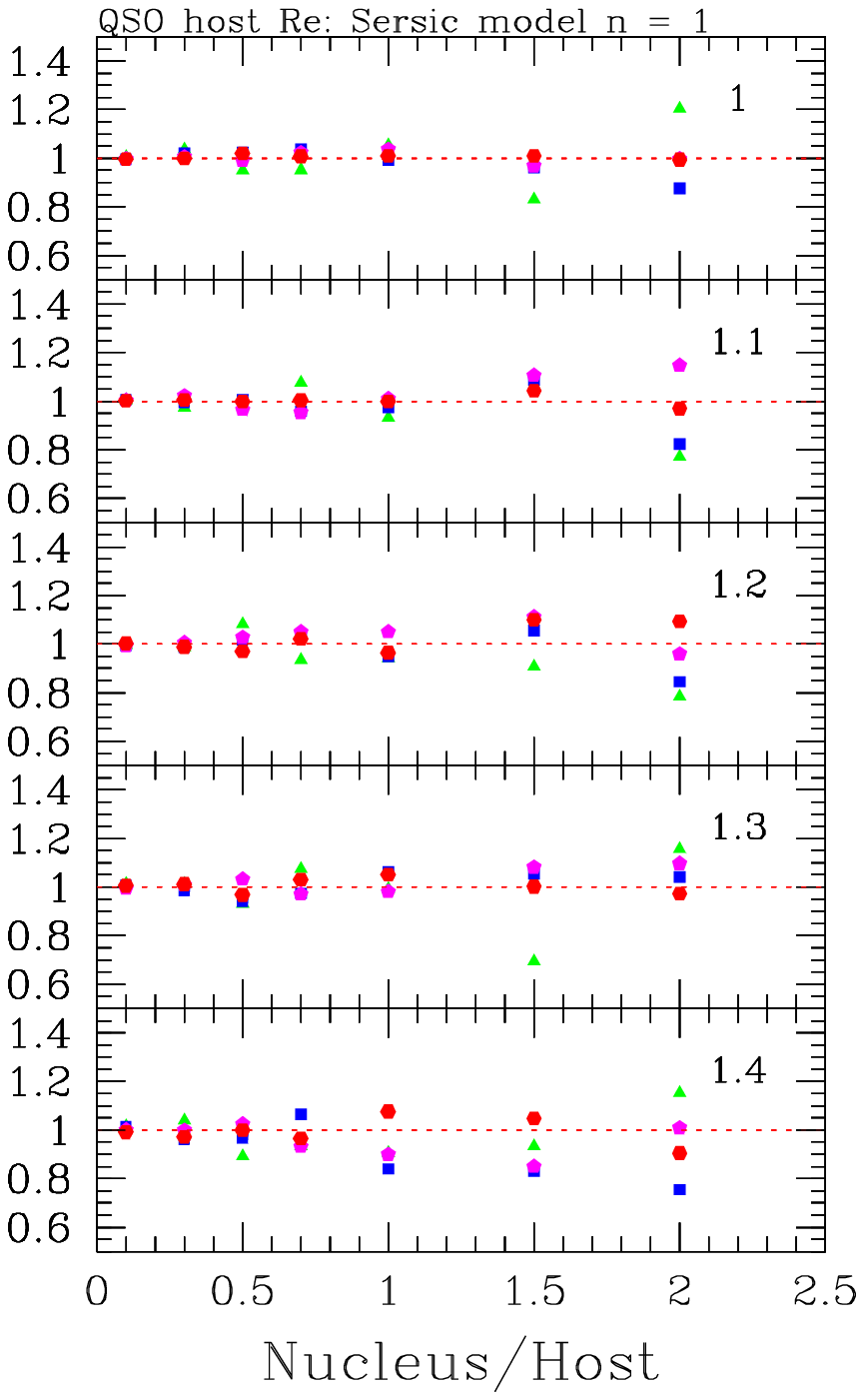}
\caption{ Same as Figure \ref{fig:mock_n4} but for host galaxy with Sersic index n = 1. }
\label{fig:mock_n1}
\end{figure}

In Figure \ref{fig:mock_comb}  we show a representative example for the three adopted host galaxy models 
of the mock simulations together with the fit obtained following the same procedure used for the real quasars.
In Figures \ref{fig:mock_n4}, \ref{fig:mock_n25}, \ref{fig:mock_n1} we show the comparison between the measure and the true parameters of the host galaxies for the various mock simulations.

It turns out that the magnitude of the host galaxy is very well measured ($<$ 0.1 mag) for 
nucleus/galaxy flux ratio smaller than 1 and seeing better than 1.2 arcsec. 
Also the effective radius of the host galaxy is recovered within an accuracy of 20\%.
For nucleus/galaxy flux ratio greater than 1 and seeing worst than 1.2 arcsec the 
uncertainty is larger but still adequate ($\Delta m < 0-3-0.4$ ) for the results presented in this work.

\clearpage
\newpage

%--- TABLES -----------------  Table 1 ------------------------------------

\begin{table*}
\caption{ The low redshift QSO sample from SDSS82. Only first 20 items are  shown. 
The complete table is available in electronic format.}

\begin{tabular}{|r|l|r|r|r|r|l|r|r|r|}
\hline
  \multicolumn{1}{|c|}{Nr$^a$} &
  \multicolumn{1}{c|}{SDSS} &
  \multicolumn{1}{c|}{RAJ2000} &
  \multicolumn{1}{c|}{DEJ2000} &
  \multicolumn{1}{c|}{z} &
  \multicolumn{1}{c|}{i$^b$} &
  \multicolumn{1}{c|}{M$_i$} &
  \multicolumn{1}{c|}{Nr.exp} &
  \multicolumn{1}{c|}{ext(i)$^c$} &
  \multicolumn{1}{c|}{psf} \\
& & & & & mag & mag & & mag & arcsecs  \\
\hline
  1 & 203657.28+000144.3 & 309.23868 & 0.02899 & 0.4412 & 19.902 & -22.02 & 8 & 0.18 & 1.27\\
  2 & 203746.78+001837.2 & 309.44492 & 0.31035 & 0.4503 & 19.58 & -22.4 & 35 & 0.16 & 1.17\\
  3 & 203905.23-005004.9 & 309.7718 & -0.83471 & 0.427 & 19.72 & -22.12 & 34 & 0.19 & 1.14\\
  4 & 204153.51+002909.8 & 310.47298 & 0.48607 & 0.3969 & 18.554 & -23.1 & 30 & 0.19 & 1.16\\
  5 & 204340.03+002853.4 & 310.91681 & 0.48151 & 0.3166 & 18.974 & -22.1 & 32 & 0.15 & 1.17\\
  6 & 204433.61+005035.5 & 311.14007 & 0.84322 & 0.4854 & 19.599 & -22.58 & 32 & 0.18 & 1.38\\
  7 & 204527.70-003236.2 & 311.36543 & -0.5434 & 0.2969 & 18.544 & -22.37 & 37 & 0.19 & 1.2\\
  8 & 204621.29+004427.8 & 311.58874 & 0.74106 & 0.4003 & 19.37 & -22.3 & 40 & 0.23 & 1.28\\
  9 & 204626.10+002337.7 & 311.60877 & 0.39381 & 0.3323 & 17.815 & -23.38 & 38 & 0.2 & 1.21\\
  10 & 204635.37+001351.7 & 311.64741 & 0.23103 & 0.4858 & 18.745 & -23.43 & 38 & 0.22 & 1.21\\
  11 & 204753.67+005324.0 & 311.97364 & 0.89001 & 0.3634 & 19.461 & -21.96 & 36 & 0.2 & 1.36\\
  12 & 204826.79+005737.7 & 312.11164 & 0.96048 & 0.4855 & 19.227 & -22.95 & 34 & 0.2 & 1.35\\
  13 & 204844.19-004721.5 & 312.18415 & -0.78931 & 0.4655 & 19.814 & -22.25 & 38 & 0.16 & 1.12\\
  14 & 204910.96+001557.2 & 312.29569 & 0.2659 & 0.3629 & 19.004 & -22.42 & 40 & 0.2 & 1.19\\
  15 & 204936.47+005004.6 & 312.40197 & 0.83462 & 0.4751 & 19.952 & -22.17 & 38 & 0.21 & 1.25\\
  16 & 204956.61-001201.7 & 312.4859 & -0.20048 & 0.3693 & 17.822 & -23.64 & 38 & 0.18 & 1.17\\
  17 & 205050.78+001159.7 & 312.71159 & 0.19992 & 0.3089 & 19.024 & -21.99 & 37 & 0.22 & 1.09\\
  18 & 205105.02-003302.7 & 312.77092 & -0.55077 & 0.2999 & 18.957 & -21.98 & 38 & 0.25 & 1.19\\
  19 & 205212.28-002645.2 & 313.0512 & -0.44589 & 0.2675 & 18.356 & -22.3 & 36 & 0.25 & 1.21\\
  20 & 205352.03-001601.5 & 313.46682 & -0.2671 & 0.3626 & 18.921 & -22.5 & 32 & 0.21 & 1.08\\
    \hline
    \end{tabular}
    \begin{list}{}
 \item   $^{(a)}$ An asterisk indicate radio loud QSO from FIRST \citep{becker97,becker12}  \\
    $^{(b)}$ psf magnitude for filter $i$ from SDSS-DR7 \\     
    $^{(c)}$ Extinction from SDSS DR7 \citep{abazajian09}  \\
        \end{list}
        \label{tab1}
\end{table*}

% --- Table 2 ---
\clearpage
\newpage
\begin{table*}
\caption{ The properties of QSO host galaxies. Only first 30 items are  shown. 
The complete table is available in electronic format.}
\begin{tabular}{|r|l|r|r|r|r|r|r|l|l|}
\hline
  \multicolumn{1}{|c|}{Nr} &
  \multicolumn{1}{c|}{SDSS} &
  \multicolumn{1}{c|}{z} &
  \multicolumn{1}{c|}{nucleus} &
  \multicolumn{1}{c|}{host} &
  \multicolumn{1}{c|}{$R_e$} &
  \multicolumn{1}{c|}{$Chi^2_{ser}$} &
  \multicolumn{1}{c|}{$Chi^2_{PSF}$} &
  \multicolumn{1}{c|}{C} &
  \multicolumn{1}{c|}{T} \\
& & & mag  & mag & & & & &   \\
\hline
  1 & 203657.28+000144.3 & 0.4412 & 20.37 & 19.49$\pm$0.15 &  0.86$\pm$0.2  & 22.27 & 46.21 & r & c\\
  2 & 203746.78+001837.2 & 0.4503 & 19.84 &  $>$19.8  & ... & 12.23 & 13.21 & u & f\\
  3 & 203905.23-005004.9 & 0.427 & 19.54 & 20.86$\pm$0.2 &  (0.42$\pm$0.11)  & 11.58 & 12.68 & m & n\\
  4 & 204153.51+002909.8 & 0.3969 & 18.97 & 18.55$\pm$0.15 & 1.7$\pm$0.17 & 6.64 & 27.06 & r & c\\
  5 & 204340.03+002853.4 & 0.3166 & 19.0 & 20.38$\pm$0.15 & 0.77$\pm$0.19 & 14.35 & 20.0 & r & f\\
  6 & 204433.61+005035.5 & 0.4854 & 19.91 & 19.57$\pm$0.25 & (2.77$\pm$0.39) & 2.53 & 6.15 & m & c\\
  7 & 204527.70-003236.2 & 0.2969 & 18.81 &  $>$18.8  & ... & 9.44 & 10.78 & u & f\\
  8 & 204621.29+004427.8 & 0.4003 & 20.27 & 21.04$\pm$0.1 & 1.03$\pm$0.23 & 14.47 & 16.33 & r & n\\
  9 & 204626.10+002337.7 & 0.3323 & 17.93 & 19.55$\pm$0.25 & (0.87$\pm$0.21) & 11.64 & 17.23 & m & c\\
  10 & 204635.37+001351.7 & 0.4858 & 18.97 & 19.34$\pm$0.15 & 1.0$\pm$0.23 & 5.83 & 10.26 & r & c\\
  11 & 204753.67+005324.0 & 0.3634 & 19.67 & 19.56$\pm$0.15 & 1.42$\pm$0.29 & 7.62 & 19.34 & r & c\\
  12 & 204826.79+005737.7 & 0.4855 & 19.59 & 19.52$\pm$0.15 & 0.68$\pm$0.17 & 15.83 & 23.14 & r & c\\
  13 & 204844.19-004721.5 & 0.4655 & 20.05 & 19.34$\pm$0.1 & 1.39$\pm$0.28 & 15.48 & 38.89 & r & n\\
  14 & 204910.96+001557.2 & 0.3629 & 18.93 &  $>$19.24   & ... & 12.52 & 14.24 & u & c\\
  15 & 204936.47+005004.6 & 0.4751 & 20.35 & 20.35$\pm$0.25 & (0.55$\pm$0.14) & 14.48 & 17.03 & m & f\\
  16 & 204956.61-001201.7 & 0.3693 & 17.57 & 19.45$\pm$0.15 & 1.27$\pm$0.27 & 16.02 & 22.92 & r & c\\
  17 & 205050.78+001159.7 & 0.3089 & 19.45 & 18.12$\pm$0.1 & 1.54$\pm$0.15 & 10.86 & 114.77 & r & n\\
  18 & 205105.02-003302.7 & 0.2999 & 19.29 & 18.48$\pm$0.15 & 2.04$\pm$0.2 & 3.92 & 36.31 & r & c\\
  19 & 205212.28-002645.2 & 0.2675 & 18.43 & 18.24$\pm$0.15 & 2.42$\pm$0.24 & 15.24 & 81.7 & r & c\\
  20 & 205352.03-001601.5 & 0.3626 & 19.1 & 20.32$\pm$0.15 & 0.61$\pm$0.16 & 12.54 & 15.39 & r & f\\
\hline
     \end{tabular}
     \begin{list}{}
       \item Notes: nucleus and host  magnitudes are given in the SDSS i band. \\ 
         C is the class type: r=resolved object, 
     u=unresolved ,m=marginally resolved, x=discarded object \\ 
     T is the type of environment n=no feature visible, 
f=features visible, c=companions visible in the nearby field \\
\end{list}
 \label{tab:results}  
\end{table*}

% --- Table 3 -----------------------------------------------------------------------
\clearpage
\newpage

\begin{table*}
\caption{ The properties of QSO host galaxies. Only first 20 items are  shown. 
The complete table is available in electronic format.}
\begin{tabular}{|r|l|r|r|r|r|r|l|r|l|}
\hline
  \multicolumn{1}{|c|}{ID} &
  \multicolumn{1}{c|}{Object name} &
  \multicolumn{1}{c|}{z} &
  \multicolumn{1}{c|}{k-cor} &
  \multicolumn{1}{c|}{$M_{R}(nuc)$} &
  \multicolumn{1}{c|}{$M_{R}(host)$} &
  \multicolumn{1}{c|}{$R_e$} &
  \multicolumn{1}{c|}{C} &
  \multicolumn{1}{|c|}{$e$} &
  \multicolumn{1}{l|}{Morph type} \\
& & & mag & mag & mag & kpc & & & \\
\hline
  1 & 203657.28+000144.3 & 0.441 & 0.2    & -22.12 &      -23.04   & 6.88 & r    &      0.43 & e\\         
  2 & 203746.78+001837.2 & 0.45 & 0.19    & -22.67 &   $>$-22.76   & ... & u  &      .... & ...\\      
  3 & 203905.23-005004.9 & 0.427 & 0.22   & -22.89 &      -21.57   & 3.26 & m   &      .... & ...\\      
  4 & 204153.51+002909.8 & 0.397 & 0.27   & -23.29 &      -23.65   & 12.73 & r  &      0.15 & d\\         
  5 & 204340.03+002853.4 & 0.317 & 0.36   & -22.60 &      -21.12   & 4.99 & r   &      0.12 & n\\         
  6 & 204433.61+005035.5 & 0.485 & 0.14   & -22.76 &      -23.26   & 23.3 & m   &      .... & ...\\      
  7 & 204527.70-003236.2 & 0.297 & 0.38   & -22.54 &   $>$-22.57   & ... & u &      .... & ...\\      
  8 & 204621.29+004427.8 & 0.4 & 0.27     & -22.05 &      -21.22   & 7.75 & r     &      0.35 & n\\         
  9 & 204626.10+002337.7 & 0.332 & 0.34   & -23.88 &      -22.14   & 5.8 & m     &      .... & ...\\      
  10 & 204635.37+001351.7 & 0.486 & 0.14  & -23.75 &      -23.54   & 8.42 & r  &      0.19 & n\\         
  11 & 204753.67+005324.0 & 0.363 & 0.31  & -22.38 &      -22.37   & 10.06 & r &      0.51 & d\\         
  12 & 204826.79+005737.7 & 0.486 & 0.14  & -23.10 &      -23.33   & 5.69 & r   &      0.3 & n\\          
  13 & 204844.19-004721.5 & 0.466 & 0.17  & -22.53 &      -23.33   & 11.44 & r &      0.14 & n\\         
  14 & 204910.96+001557.2 & 0.363 & 0.31  & -23.12 &   $>$-22.70   & ... & u &      .... & ...\\      
  15 & 204936.47+005004.6 & 0.475 & 0.16  & -22.31 &      -22.44   & 4.59 & m  &      .... & ...\\      
  16 & 204956.61-001201.7 & 0.369 & 0.31  & -24.51 &      -22.51   & 9.08 & r   &      0.34 & n\\         
  17 & 205050.78+001159.7 & 0.309 & 0.37  & -22.10 &      -23.38   & 9.81 & r  &      0.17 & d\\         
  18 & 205105.02-003302.7 & 0.3 & 0.38    & -22.15 &      -22.97   & 12.73 & r    &      0.42 & d\\         
  19 & 205212.28-002645.2 & 0.268 & 0.4   & -22.62 &      -22.92   & 13.94 & r  &      0.3 & d\\          
  20 & 205352.03-001601.5 & 0.363 & 0.31  & -22.97 &      -21.64   & 4.29 & r  &      0.0 & n\\     
\hline
\end{tabular}
\begin{list}{}
      \item Note: M$_R$ nucleus and host magnitudes are in the R band k-corrected. \\ 
         R$_e$  is effective radius in kpc; 
         C is the e class type: r=resolved object, 
      u=unresolved ,m=marginally resolved, x=discarded object \\ 
     e is the ellipticity of the host galaxy 
      and Morph type is the following:  e=elliptical dominant, d=disk dominant, n=not classifiable.\\
\end{list}
 \label{tab:abs}    
\end{table*}

\clearpage

% --- REFERENCES -------------------
%\newpage
%\null

\begin{figure*}
\centering
\includegraphics[width=2.0\columnwidth]{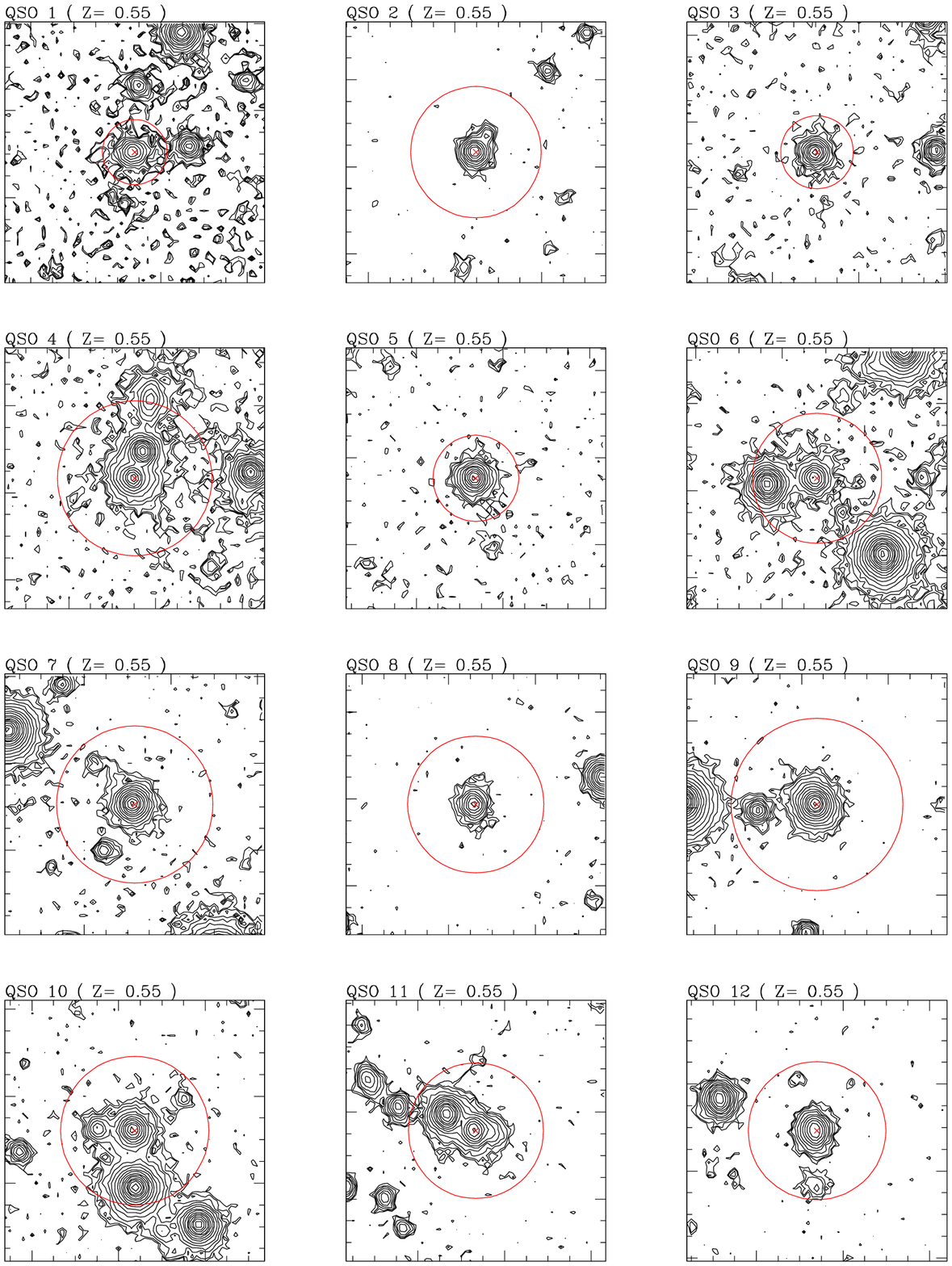}
\caption{Contour plot of the first 12 targets (the whole figure is available as online supplementary material). The QSO is at the center of each panel and marked with a red square.
The field of view in each box is 24 arcsec across. The red central circle represents the 
region of the fit. Masked out objects are not shown here (see example in Fig. \ref{fig:psf})}
\label{fig:fit1}
\end{figure*}

\begin{figure*}
\centering
\includegraphics[width=2.0\columnwidth]{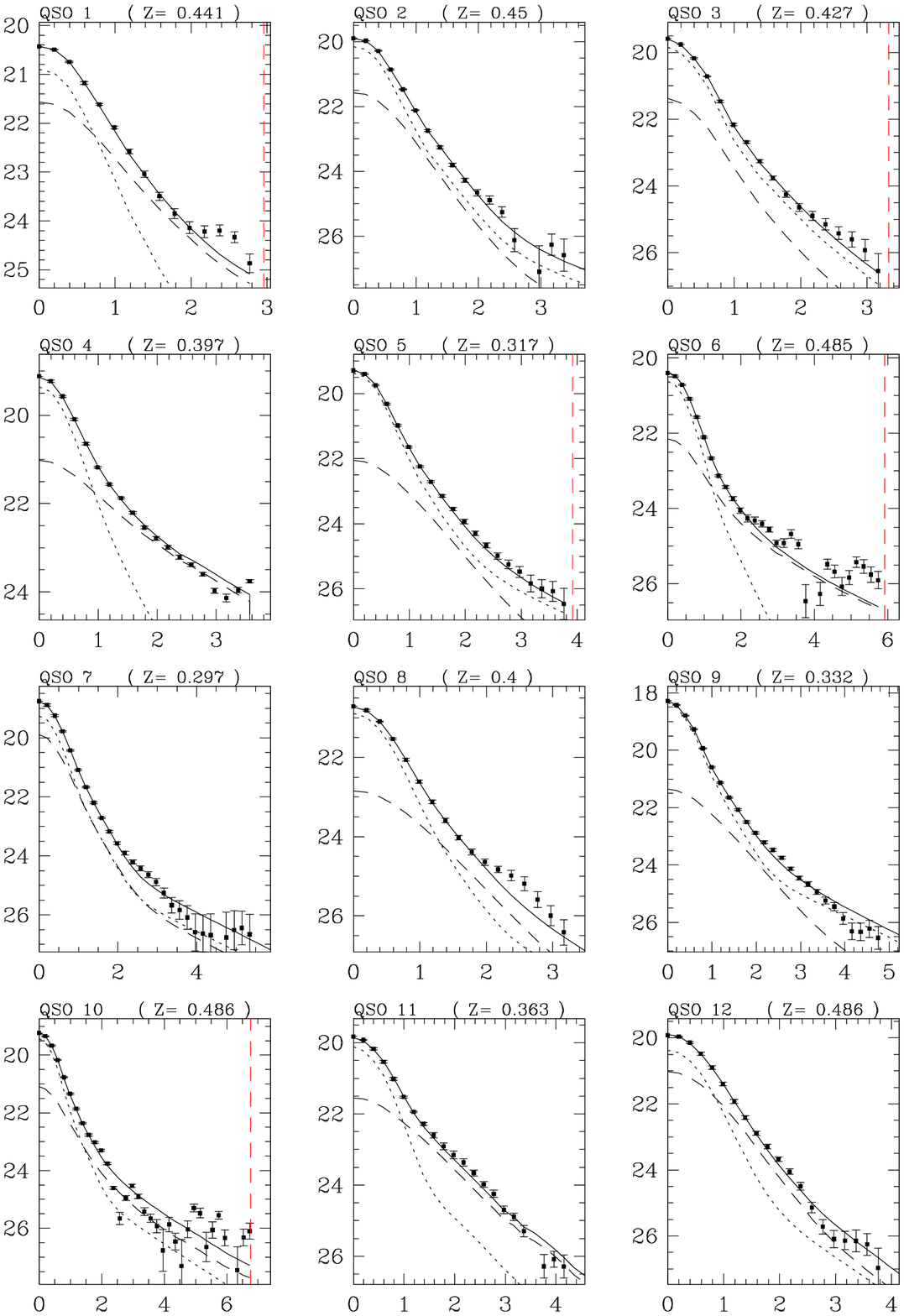}
\label{fig:fit2}
\caption{Fit of the first 12 targets (the whole figure is available as online supplementary material). Observed radial brightness profile (filled squares) compared with the model 
fit (solid line) with the two components: scaled PSF (dotted line) and galaxy 
model (dashed line) convolved with the proper PSF.  The vertical dashed red line represents the limit of the region for the fit of the data.}
\end{figure*}

\section*{Acknowledgments}

We are grateful to the anonymous referee for detailed comments that led to improve 
the presentation of these  results. 
Funding for the SDSS and SDSS-II has been provided by the Alfred P. Sloan Foundation, the Participating Institutions, the
National Science Foundation, the U.S. Department of Energy, the National Aeronautics and Space Administration, the Japanese
Monbukagakusho, the Max Planck Society, and the Higher Education Funding Council for England. The SDSS Web Site is
http://www.sdss.org/.

The SDSS is managed by the Astrophysical Research Consortium for the Participating Institutions. The Participating
Institutions are the American Museum of Natural History, Astrophysical Institute Potsdam, University of Basel, University
of Cambridge, Case Western Reserve University, University of Chicago, Drexel University, Fermilab, the Institute for
Advanced Study, the Japan Participation Group, Johns Hopkins University, the Joint Institute for Nuclear Astrophysics, the
Kavli Institute for Particle Astrophysics and Cosmology, the Korean Scientist Group, the Chinese Academy of Sciences (
LAMOST), Los Alamos National Laboratory, the Max-Planck-Institute for Astronomy (MPIA), the Max-Planck-Institute for
Astrophysics (MPA), New Mexico State University, Ohio State University, University of Pittsburgh, University of
Portsmouth, Princeton University, the United States Naval Observatory, and the University of Washington.

\end{document}